\pdfoutput=1
\documentclass[12pt,letterpaper]{article}
\usepackage{putex}
\usepackage{graphicx}
\usepackage{latexsym,amsmath,amsfonts,amssymb,amsthm}
\usepackage{bbm}
\usepackage{cite}
\usepackage{indentfirst}
\usepackage{booktabs,array}
\usepackage{placeins}
\usepackage{mathtools}
\usepackage{cancel}
\usepackage[mode=image|tex]{standalone}
\usepackage[dvipsnames]{xcolor}
\usepackage{tikz}
\usepackage{tikz-cd}
\usepackage{adjustbox}
\usepackage{enumitem}
\usepackage{setspace}
\usepackage{mathdots}
\usetikzlibrary{decorations.markings}
\usepackage{graphicx}
\usepackage[margin=10pt,font=small,labelfont=bf]{caption}
\usepackage{subcaption}
\usepackage{xcolor}
\usepackage{float}
\theoremstyle{definition}

\usepackage{microtype}
\usepackage{hyperref}
\hypersetup{unicode}
\hypersetup{linktoc = all}
\hypersetup{pdfborderstyle={/S/U/W 0.5}}
\hypersetup{linkbordercolor = gray}
\hypersetup{bookmarksnumbered}
\usepackage[T1]{fontenc}
\usepackage[utf8]{inputenc}
\usepackage{lmodern}

\newcommand{\startofsentenceTF}[2]{\ifnum\spacefactor>1000 #1\else #2\fi}

\setlist{itemsep=2pt plus 1pt minus 1pt, topsep=2pt plus 1pt minus 1pt}

\renewenvironment{thebibliography}[1]{%
\begin{oldthebibliography}{#1}%
\setlength\itemsep{-2mm}%
}%
{%
\end{oldthebibliography}%
}

\newcommand{\ie}{\textsl{i.e.\@}}

\newcommand{\etc}{\textsl{etc.\@}}

\numberwithin{equation}{section}

\newcommand{\nn}{\nonumber}

\newcommand{\unit}{\mathbbm{1}}

\DeclareMathOperator{\Tr}{Tr}

\DeclareMathOperator{\re}{\mathbb{R}e}
\DeclareMathOperator{\im}{\mathbb{I}m}

\DeclareMathOperator*{\SumInt}{%
\mathchoice%
  {\ooalign{$\displaystyle\sum$\cr\hidewidth$\displaystyle\int$\hidewidth\cr}}
  {\ooalign{\raisebox{.14\height}{\scalebox{.7}{$\textstyle\sum$}}\cr\hidewidth$\textstyle\int$\hidewidth\cr}}
  {\ooalign{\raisebox{.2\height}{\scalebox{.6}{$\scriptstyle\sum$}}\cr$\scriptstyle\int$\cr}}
  {\ooalign{\raisebox{.2\height}{\scalebox{.6}{$\scriptstyle\sum$}}\cr$\scriptstyle\int$\cr}}
}

\newcommand\qq{\mathbbmtt{Q}}

\begin{document}


\title{\begin{LARGE}
Chiral Algebras, Localization and Surface Defects
\end{LARGE}\newline}

\authors{Yiwen Pan$^1$ and Wolfger Peelaers$^2$
\medskip\medskip\medskip\medskip
 }

\institution{UU}{${}^1$
Department of Physics and Astronomy, Uppsala University, \cr
$\;\,$ Box 516, SE-75120 Uppsala, Sweden}
\institution{NHETC}{${}^2$
New High Energy Theory Center, Rutgers University,  \cr
$\;\,$ Piscataway, NJ 08854, USA}

\abstract{\begin{onehalfspace}{
Four-dimensional $\mathcal N=2$ superconformal quantum field theories contain a subsector carrying the structure of a chiral algebra. Using localization techniques, we show for the free hypermultiplet that this structure can be accessed directly from the path integral on the four-sphere. We extend the localization computation to include supersymmetric surface defects described by a generic 4d/2d coupled system. The presence of a defect corresponds to considering a module of the chiral algebra: our results provide a calculational window into its structure constants.
}\end{onehalfspace}}

\preprint{UUITP-34/17}
\setcounter{page}{0}
\maketitle


{
\setcounter{tocdepth}{2}
\setlength\parskip{-0.7mm}
\tableofcontents
}


\section{Introduction}
Four-dimensional $\mathcal N=2$ superconformal quantum field theories (SCFTs) contain a subsector of local operators whose operator product algebra is described by a chiral algebra \cite{Beem:2013sza}. The chiral algebra repackages and organizes an infinite amount of conformal data in a tightly constrained structure. As a result, it is an efficient tool to obtain new results on various aspects of the above-lying SCFT. For example, one can extract novel unitarity bounds on central charges \cite{Beem:2013sza,Liendo:2015ofa,Lemos:2015orc,Cornagliotto:2017dup}, get a handle on Higgs branches and their relations \cite{Lemos:2014lua,Beem:2017ooy}, find expressions for superconformal indices \cite{Buican:2015ina,Cordova:2015nma,Buican:2015tda,Song:2015wta,Cordova:2016uwk,Cordova:2017ohl,Buican:2017uka,Neitzke:2017cxz} and obtain modular differential equations they must satisfy \cite{Arakawa:2016hkg,Beem:2017ooy,Beem:WIP1}, \etc{} \textit{Vice versa}, the image of the map from four-dimensional $\mathcal N=2$ SCFTs to chiral algebras defines a set of vertex operator algebras of mathematical interest \cite{Arakawa:2015jya,Arakawa:2016hkg}. For example, the image of theories of class $\mathcal S$ has the structure of a topological quantum field theory valued in chiral algebras \cite{Beem:2014rza}. 

The SCFT/chiral algebra correspondence was constructed algebraically in \cite{Beem:2013sza}: the relevant subsector of local operators was isolated by passing to the cohomology of either one of two well-chosen nilpotent supercharge $\qq_i,\, i=1,2$, and it was shown that the algebra of cohomology classes, obtained by reducing the operator product algebra, is isomorphic to a chiral algebra. The chiral algebra depends meromorphically on the complex coordinates of a plane that is singled out by the choice of $\qq_i$. The first goal of this paper is to show that the chiral algebra structure can be carved out directly from the path integral. We restrict attention to the example of the free hypermultiplet, which already showcases all salient features and serves as a proof of principle. Concretely, we map the theory to the round four-sphere, and employ supersymmetric localization techniques with respect to the supercharge $\mathcal Q = \qq_1+\qq_2$ to argue that the theory can be localized to a quantum field theory on a two-sphere.\footnote{This computation is similar in spirit to \cite{Pestun:2009nn}. See also \cite{Dedushenko:2016jxl} for an analogous computation on the three-sphere. } This quantum field theory is precisely the chiral algebra associated with the free hypermultiplet, namely the symplectic boson pair. The localization computation is compatible with the insertion of operators in the subsector mentioned above. Schematically,
\begin{equation}
\int [D\Phi_{\text{HM}}] \ \mathcal O\  e^{-S_{\text{HM}}[\Phi_{\text{HM}}]} = \int [DQ D\widetilde Q] \ \mathcal O \ e^{-S_{{\text{SB}}}[Q,\widetilde Q]}\;.
\end{equation}
Here $\Phi_{\text{HM}}$ collectively denotes all fields in the hypermultiplet and $S_{\text{HM}}[\Phi_{\text{HM}}]$ is the hypermultiplet action. The insertion $\mathcal O$ represents any collection of local operators belonging to the subsector and whose correlator we wish to compute. On the right hand side, $Q$ and $\widetilde Q$ are the symplectic boson pair with action $S_{{\text{SB}}}[Q,\widetilde Q]$ on the two-sphere.

The representation theory of the chiral algebras associated with $\mathcal N=2$ SCFTs can be probed by inserting surface defects in the four-dimensional theory \cite{Beem:2013sza,Cordova:2017mhb,Beem:WIP2}. Indeed, it is easy to verify that a defect preserving $\mathcal N=(2,2)$ supersymmetry on its worlvolume can be embedded perpendicular to the plane in which the chiral algebra lives in such a way that it is compatible with the nilpotent supercharges $\qq_i$. Its insertion corresponds to considering a module of the chiral algebra. Our localization computation can be extended likewise to include defects wrapping an orthogonal two-sphere. We consider a large class of defects described by coupling an arbitrary two-dimensional quantum field theory $\mathcal T_{\text{2d}}$ to the free hypermultiplet via a twisted superpotential coupling.\footnote{The embedding of the two-dimensional $\mathcal N=(2,2)$ symmetry algebra in the four-dimensional $\mathcal N=2$ algebra is such that a hypermultiplet decomposes into a pair of twisted chiral multiplets (and their conjugates).}. We present in detail the extension of the localization computation to these 4d/2d coupled systems. Our result for the partition function of the 4d/2d coupled system is
\begin{equation}
Z_{\text{4d/2d}} = \SumInt [d\phi_{\text{2d}}]\ Z_{\text{1-loop}}(\phi_{\text{2d}}) \int [DQ D\widetilde Q]\ e^{-S_{\text{SB}}[Q,\widetilde Q]} \ e^{- 4\pi ir \big[\widetilde W(Q(\text{NP}),\widetilde Q(\text{NP}),\phi_{2d}^{\text{tc}}) + c.c. \big]}\;.
\end{equation}
Here $\phi_{\text{2d}}$ collectively denotes the localization locus of the two-dimensional theory $\mathcal T_{\text{2d}}$, and $Z_{\text{1-loop}}(\phi_{\text{2d}})$ is the one-loop determinant of quadratic fluctuations of the two-dimensional theory. In the twisted superpotential, $\widetilde W$, the quantum fields $Q$ and $\widetilde Q$ are pinned at the north pole and we collectively wrote $\phi_{2d}^{\text{tc}}$ for the bottom components of the twisted chiral multiplets of the two-dimensional theory. The latter are set to their constant BPS profile (which is integrated/summed over). In the complex conjugate term, the symplectic boson pair is located at the south pole. This result can be further enriched by operator insertions. Away from the north and south poles, the computation is compatible with the insertion of operators in the subsector as above, while at the poles one can additionally insert native two-dimensional twisted chiral fields. This distinction of allowed insertions between the bulk of the two-sphere and the poles precisely reflect the structure of the module.

While information about the space of states constituting the module can be gained via, for example, the superconformal index, the structure constants of the module have been inaccessible so far. Our localization computation opens a computational window to compute these coefficients and to analyze their dependence on coupling constants.

The paper is organized as follows. In section \ref{section:review-chiral-algebra} we briefly review the salient features of the SCFT/chiral algebra correspondence. In section \ref{section:setup} we prepare the localization computation by placing the theory on the four-sphere and selecting the localizing supercharge. The main results of the paper are in sections \ref{section:freeHM} and \ref{section:defects}, in which we present the localization computation of the free hypermultiplet to the symplectic boson, and its extension to include surface defects described by 4d/2d coupled systems. Two appendices contain some technical details and further results.


\section{Review of SCFT/chiral algebra correspondence}\label{section:review-chiral-algebra}
In this section, we briefly review the correspondence between four-dimensional $\mathcal N=2$ superconformal field theories and chiral algebras. We refer the readers to the original paper \cite{Beem:2013sza} for all details. We also explain how the inclusion of a superconformal surface defect in the four-dimensional SCFT naturally leads one to consider modules of the chiral algebra \cite{Beem:2013sza,Cordova:2017mhb,Beem:WIP2}. 

As for any conformal field theory, the conformal data of a four-dimensional $\mathcal N=2$ SCFT fully determine correlation functions of finitely many local operators. These data comprise the spectrum of local operators, organized in representations of the superconformal algebra and any additional global symmetry algebras, and the three-point couplings or operator product expansion (OPE) coefficients, one for every three local operators. Naturally, the organization of local operators in multiplets implies relations among various OPE coefficients. The conformal data satisfy the conformal bootstrap constraints, which state that the operator product algebra is associative. The conformal bootstrap program aims to reverse the logic by trying to extract useful information from these constraints, and ideally solve for the conformal data, given a minimal amount of information on the (S)CFT. It has been successfully implemented for rational conformal field theories in two dimensions, in large part thanks to their enlarged Virasoro symmetry, but is not easily generalizable analytically away from rationality or for higher-dimensional theories. In \cite{Beem:2013sza}, however, it was found that for four-dimensional $\mathcal N=2$ SCFTs a solvable truncation can be obtained by considering instead of the algebra of local operators the algebra of \textit{cohomology classes} with respect to a cleverly chosen nilpotent supercharge.\footnote{See \cite{Beem:2014kka,Chester:2014mea,Beem:2016cbd} for similar constructions in three and six dimensions.} This algebra was shown to be a chiral algebra.

In the $\mathcal N=2$ superconformal algebra, one can define two nilpotent supercharges, denoted as $\qq_i$, $i=1,2$, which each take the schematic form of a sum of a Poincar\'e supercharge and a superconformal supercharge.\footnote{\label{Qforcohomology}Denoting the Poincar\'e supercharges of the four-dimensional $\mathcal N=2$ superconformal algebra as $\mathcal Q_{I\alpha}, \widetilde {\mathcal Q}^I_{\dot\alpha}$ and the conformal supercharges as $\mathcal S^{I\alpha}, \widetilde {\mathcal S}^{\dot\alpha}_I$, we take, following \cite{Beem:2013sza}, $\qq_1 = \mathcal Q_{1-} + \widetilde{ \mathcal S}^{\dot -}_2$ and $\qq_2 = \mathcal S^{1-} - \widetilde{\mathcal Q}_{\dot -}^2$. Here $I$ is an $SU(2)_R$ index and $\alpha,\dot\alpha$ are the standard spinor indices.} Their anticommutator is given by
\begin{equation}
\{\qq_1,\qq_2 \} = -\mathcal M^{\bot} - r\;,
\end{equation}
where $r$ denotes the $U(1)_r$ charge and $\mathcal M^{\bot}$ is the generator of rotations in the $(x_1,x_2)$-plane. The latter can be expressed in terms of the usual rotation generators $(j_1,j_2)$ as $\mathcal M^{\bot}=j_1-j_2$. The cohomologies of $\qq_i$ can be shown to be isomorphic and can be easily characterized in two steps. First, the harmonic representatives of cohomology classes at the origin are characterized by the conditions
\begin{equation}
\frac{1}{2}(E-j_1-j_2)-R=0\;, \qquad r+(j_1-j_2)=0\;,
\end{equation}
which contain the conformal dimension $E$ and the $SU(2)_R$ spin $R$. Operators satisfying these conditions were called Schur operators in \cite{Beem:2013sza} as they satisfy the shortening conditions defining operators contributing to the Schur limit of the superconformal index \cite{Gadde:2011uv}. Unitarity of the four-dimensional SCFT demands that Schur operators necessarily are $SU(2)_R$ highest weight states, and moreover that their rotational quantum numbers $j_1$ and $j_2$ are maximal. A Schur operator can thus be schematically represented as $\mathcal O_{(11\ldots1),+\ldots +\dot + \ldots \dot +}$, where the first set of (symmetrized) indices are $SU(2)_R$ indices, and the other indices are the usual Lorentz indices. We will sometimes suppress the Lorentz indices in what follows. Second, one can move the operator away from the origin, while remaining inside the cohomology of $\qq_i$, by employing $\qq_i$-closed translation operators. It turns out that one can only translate the operators in this manner in the plane kept fixed pointwise by $\mathcal M^{\bot}$, which we will call the chiral algebra plane. Introducing complex coordinates $(z,\bar z)$ in this plane, the $z$-translation operator $\mathcal P_z$ is $\qq_i$-closed, while a twisted $\bar z$-translation is $\qq_i$-exact.\footnote{Note that in spinorial notation $x^{+\dot+} = z$ and $x^{-\dot -}=\bar z$.} The latter is given concretely as $\mathcal P_{\bar z} + R^-$, where $R^-$ is the $SU(2)_R$ lowering operator. One then finds
\begin{align}
\mathcal O(z,\bar z) &= e^{z\mathcal P_z + \bar z(\mathcal P_{\bar z} + R^-)}\  \mathcal O_{(11\ldots1)}(0)\  e^{-z\mathcal P_z - \bar z(\mathcal P_{\bar z} + R^-)}  \\
&= u^{I_1}(\bar z) \ldots u^{I_k}(\bar z) \mathcal O_{(I_1 I_2\ldots I_k)}(z,\bar z)\;, \qquad \text{with}\qquad u^I(\bar z) = (1,\  \bar z)\;.\label{twistedtranslation}
\end{align}
As the $\bar z$ dependence is generated by a $\qq_i$-exact operator, it drops out in cohomology: $[\mathcal O(z,\bar z)]_{\qq_i} = \mathcal O(z)$. One can argue that the algebra of cohomology classes thus obtained defines a chiral algebra.

The resulting chiral algebras exhibit many beautiful, general properties \cite{Beem:2013sza}:
\begin{itemize}
\item It features Virasoro symmetry with central charge $c_{2d} = -12 c_{4d}$, where $c_{4d}$ is the (four-dimensional) Weyl anomaly associated to the square of the Weyl tensor. It is noteworthy that four-dimensional unitarity implies that the chiral algebra is necessarily non-unitary.
\item Four-dimensional flavor symmetries manifest themselves as affine symmetries of the chiral algebra. The level $k_{2d}$ of the current algebra is universally determined in terms of the flavor central charge $k_{4d}$, which controls the canonically normalized flavor current two-point function, as $k_{2d} = -\frac{1}{2} k_{4d}$.
\item Exactly marginal gauging in the four-dimensional SCFT is concisely captured by a BRST reduction. Non-renormalization theorems guarantee that the resulting chiral algebra is independent of the exactly marginal coupling.\footnote{See section 3.4.2 of \cite{Beem:2013sza}.}
\item Each generator of the Higgs branch chiral ring gives rise to a strong generator of the chiral algebra,\footnote{In fact, each generator of the Hall-Littlewood chiral ring, introduced in \cite{Beem:2013sza}, descends to a strong generator.} while Higgs branch chiral ring relations translate into null relations. Moreover, the Higgs branch can be extracted from the chiral algebra as its associated variety \cite{Beem:2017ooy}. 
\item The vacuum character of the chiral algebra is computed by the Schur limit of the superconformal index.
\end{itemize} 

The chiral algebras associated to free $\mathcal N=2$ SCFTs, \ie{}, the free hypermultiplet and the free vectormultiplet, are easily found to be a symplectic boson pair and a small $(b,c)$ ghost system. As these chiral algebras are the basic building blocks of any localization computation, we briefly review their construction. The four real scalars of a free hypermultiplet are rotated by an $SU(2)_R\times SU(2)_F$ symmetry. The first factor is the R-symmetry, whose indices we denote as $I,J,\ldots$, while the second is a flavor symmetry, for which we use indices $A,B,\ldots$ Correspondingly, we write $q_{IA}=\left(\begin{smallmatrix} Q & \tilde Q \\
-\tilde Q^\dagger & Q^\dagger \end{smallmatrix}\right)$. The Schur operators in the free hypermultiplet are all words built from the elementary letters $q_{1A}$ and the derivative $\partial_{+\dot +}=\partial_{z}$. The cohomology elements are then obtained by applying the `twisted translation' of \eqref{twistedtranslation}. In particular, one finds
\begin{equation}\label{twistedtranslationfreeHM}
q_A(z) = [u^I(\bar z) q_{IA}(z,\bar z)]_{\qq_i}=[q_{1A}(z,\bar z) + \bar z\ q_{2A}(z,\bar z)]_{\qq_i}\;.
\end{equation}
The meromorphic operator product expansion (OPE) of $q_A(z)$ with itself can be easily deduced from the OPE of the four-dimensional free hypermultiplet scalars, $q_{IA}(x) q_{JB}(y)\sim \frac{\epsilon_{IJ}\epsilon_{AB}}{(x-y)^2}$, and reads
\begin{equation}
q_A(z) q_B(w) \sim \frac{\epsilon_{AB}}{z-w}\;.
\end{equation}
One readily recognizes this OPE as defining a symplectic boson pair, and one can easily convince oneself that the full algebra of cohomology classes indeed corresponds to this chiral algebra.

Similarly, the free vector multiplet contains as Schur operators all words built from the letters $\lambda_{1+},\tilde \lambda_{1\dot +}$ and $\partial_{+\dot +}$, where $\lambda_{I\alpha},\tilde \lambda_{I\dot \alpha}$ are the gaugini. The full cohomology is then obtained by `twisted translations' of these words. For the elementary letters one can define
\begin{equation}
\tilde \lambda(z) = [u^I(\bar z) \tilde \lambda_{I\dot +}(z,\bar z)]_{\qq_i}\;, \qquad \lambda(z) = [u^I(\bar z) \lambda_{I+}(z,\bar z)]_{\qq_i}\;,
\end{equation}
which can be easily shown to satisfy the OPE
\begin{equation}
\lambda(z) \tilde \lambda(w) \sim \frac{1}{(z-w)^2}\;.
\end{equation}
Upon identifying $\tilde \lambda = b$ and $\lambda = \partial c$, these OPEs can be recognized as defining the $(b,c)$ ghost system of type $(1,0)$, $b(z) c(w) \sim (z-w)^{-1}$, upon removing the spurious $c$ zero-mode. The resulting chiral algebra is the small $(b,c)$ ghost system. In this paper we will focus mainly focus on the free hypermultiplet and provide some comments on the free vector multiplet in appendix \ref{appendix_oneloop}.

Let us now consider a four-dimensional $\mathcal N=2$ SCFT in the presence of a half-BPS superconformal surface defect. It is straightforward to verify that if the defect preserves $\mathcal N=(2,2)$ superconformal symmetry on its worldvolume and is transverse to the chiral algebra plane, piercing it at the origin, then it can be embedded such that both supercharges $\qq_i$ are preserved.\footnote{See subsection \ref{embeddingflatspace} for a detailed discussion of the relevant embedding.} The presence of the defect enriches the cohomology of the nilpotent supercharges $\qq_i$ at the origin, while away from the origin the twisted-translated (four-dimensional) Schur operators still make up the full cohomology. Let us denote the cohomology at the origin as $M$ and the original chiral algebra, as an algebra of modes, as $\mathcal A$. It is easy to see that in the setup at hand, $M$ is endowed with the algebraic structure of a module \cite{Beem:2013sza,Cordova:2017mhb,Beem:WIP2}:
\begin{equation}\label{algebraonmodule}
\cdot: \mathcal A \times M \rightarrow M: (\mathfrak a,m) \mapsto \mathfrak a\cdot m\;,
\end{equation}
where the action of the chiral algebra on the module is defined as follows. Let $\mathfrak a=a_{-h_a-p}$ be a mode of the twisted-translated Schur operator $a(z)$ of weight $h_{a}$, \ie{}, a Laurent coefficient of the expansion $a(z)=\sum_n a_{-h_a -n} z^{n}$. Consider the bulk-defect operator product expansion (within cohomology) of $a(z)$ with (a representative of) the cohomology class $m$. Then $\mathfrak a\cdot m$ is obtained by selecting the coefficient of $z^p$ in this OPE. One can easily verify that thanks to the associativity of the bulk-defect OPE this action satisfies the standard requirements to define a module. As always, if $e_i$ is some basis for $M$, the structure constants $\lambda_{\mathfrak ai}^{\phantom{ai}\,j}$ are defined as
\begin{equation}\label{defstructureconstants}
\mathfrak a\cdot e_i = \sum_{j} \lambda_{\mathfrak ai}^{\phantom{ai}\,j}\ e_j\;,
\end{equation}
and extended to all of $M$ by linearity.

A powerful probe to analyze the modules thus obtained and their properties is their (graded) character: it counts (with signs) the elements of $M$. This quantity can be computed as the Schur limit of the superconformal index of the four-dimensional SCFT in the presence of the surface defect, and allows to attempt to identify the module and to answer general questions like, for example, if the modules satisfy any particular representation theoretic properties, or perhaps, in the other extreme, if for any given $\mathcal N=2$ SCFT all modules of the corresponding chiral algebra can be obtained from the insertion of some surface operator.\footnote{A particularly `nice' property one could hope the module to have is a spectrum bounded from below. In examples, however, one can show this property to be violated \cite{Beem:WIP2}.} Such analysis has been initiated in \cite{Cordova:2017mhb,Beem:WIP2}. On the other hand, the character has little to say about whether the module possesses any dependence on coupling constants.\footnote{\label{footnoteparamdependence}Let us be more precise and define the dependence of a module on a parameter $\tau$. We denote by $M_\tau$ the module at the specified value of the parameter, and assume that there exists an isomorphism $\phi_{\tau'\tau}:M_\tau\rightarrow M_{\tau'}$ (for sufficiently small $|\tau-\tau'|$ and at a generic value of $\tau$) as vector spaces. Let us denote the action of the chiral algebra on the module at $\tau$ as $\cdot_\tau$. We say that the module does \textit{not} depend on the parameter $\tau$ if for any $\tau$ the following diagram commutes
\begin{center}
  \includegraphics[width=0.28\textwidth]{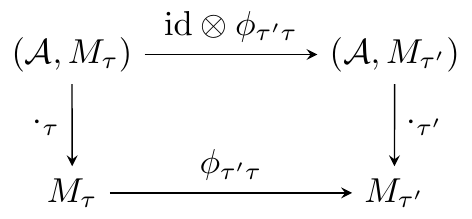}
\end{center}

Let $e_i(\tau)$ be a basis at $\tau$ and $\phi_{\tau'\tau}(e_i(\tau))=\sum_j \phi_i^{\phantom{i}\,j} e_j(\tau')$. Then commutativity of the above diagram requires that the structure constants obey
\begin{equation}
\lambda_{ai}^{\phantom{ai}\,j}(\tau) = \phi_i^{\phantom{i}\,j}\, \lambda_{aj}^{\phantom{aj}\,k}(\tau')\,(\phi^{-1})_k^{\phantom{k}\,l}\;.
\end{equation} } Unlike the dependence of the chiral algebra on exactly marginal couplings, there is no known non-renormalization theorem preventing any dependence of the module on coupling constants.\footnote{The argument used in the former case crucially uses the existence of extended superconformal symmetry and its associated Ward identities in the chiral algebra plane. The relevant supercharges in this extended algebra are not all preserved upon inserting the defect and hence the argument is not applicable in the latter case. } Using the computations in the current paper, we provide a computational tool that in principle allows one to probe this question. 


\section{Setup}\label{section:setup}
Our first goal of this paper is to rederive the chiral algebra, whose algebraic construction was reviewed in the previous section, from the path integral on the four-sphere using localization techniques. In this section we prepare ourselves for this computation by recalling the supergravity background on $S^4$ and identifying the Killing spinor associated with the supercharge $\mathcal Q$ we would like to use for the localization computation. It turns out to be most convenient to choose (the Weyl transformation of) $\qq_1+\qq_2$. 

\subsection{\texorpdfstring{$\mathcal N=2$}{N=2} supersymmetry on \texorpdfstring{$S^4$}{S4}}
The four-sphere of radius $r$ can be described as a hypersurface in $\mathbb R^5$ described by the equation
\begin{equation}
x_0^2 + x_1^2 + x_2^2 + x_3^2 + x_4^2 = r^2\;.
\end{equation}
Introducing the coordinates $\varphi, \chi \in [0, 2\pi)$, $\theta \in [0, \pi/2]$ and $\rho \in [0, \pi]$ as
\begin{equation}\label{embeddingcoordinates}
\begin{aligned}
x_0 &= r \cos\rho\;, \qquad &x_1 &= r \sin \rho \ \cos\theta\ \cos\varphi\;, \qquad &x_3 &= r \sin \rho \ \sin\theta\ \cos\chi\;, \\
& &x_2 &= r \sin \rho \ \cos\theta\ \sin\varphi\;, \qquad &x_4 &= r \sin \rho \ \sin\theta\ \sin\chi\;,
\end{aligned}
\end{equation}
its induced metric can be written in terms of the vielbeins
\begin{equation}\label{vielbeins}
e^1 = r \sin\rho \cos\theta\  d\varphi\;, \quad e^2 = r \sin\rho \sin\theta\  d\chi\;, \quad e^3 = r \sin\rho\ d\theta\;, \quad e^4 = r d\rho\;.
\end{equation}
We note that the loci at $\theta = 0$ and $\theta = \frac{\pi}{2}$ form two intersecting two-spheres, $S^2_{\theta = 0}$ and $S^2_{\theta = \frac{\pi}{2}}$, which intersect at the north pole ($\rho=0$) and south pole ($\rho=\pi$). The vielbein on these two-spheres are easily obtained from \eqref{vielbeins} and are the standard vielbeins on a two-sphere:
\begin{align}
&e^1_{S^2_{\theta=0}} = r d\rho \;,  \phantom{e^1_{S^2_{\theta=\frac{\pi}{2}}}} e^2_{S^2_{\theta=0}} = r \sin \rho\ d\varphi \;,\label{vielbeinsS20} \\ 
&e^1_{S^2_{\theta=\frac{\pi}{2}}} = r d\rho \;, \phantom{e^1_{S^2_{\theta=0}} }   e^2_{S^2_{\theta=\frac{\pi}{2}}} = r \sin \rho\ d\chi\;. \label{vielbeinsS2thetapi2}
\end{align}

An algorithmic method to place supersymmetric theories on a curved background was developed in \cite{Festuccia:2011ws}. The idea is to first couple the theory to supergravity and then to consider a rigid limit freezing the bosonic supergravity fields, both dynamical and auxiliary, to supersymmetric configurations. The equations describing such configurations are obtained by setting to zero the supergravity variations of the gravitino and other fermionic fields in the supergravity multiplet.\footnote{\label{footnote:flavorbackground}One can similarly couple the theory to a nontrivial background for flavor symmetries.} For $\mathcal N=2$ supersymmetric theories the relevant equations have been analyzed in \cite{Hama:2012bg,Pestun:2014mja}, and in particular applied to the (squashed) four-sphere.\footnote{\label{footnote:weylrescale}Note that one can straightforwardly place a superconformal quantum field theory on a conformally flat space by performing a Weyl transformation. This approach was taken in the original paper \cite{Pestun:2007rz}. Mass deformations of the theory can be obtained as in footnote \ref{footnote:flavorbackground}. Squashing the four-sphere, however, requires the full machinery of \cite{Festuccia:2011ws}.}

On the four-sphere, the result of the above analysis (or the shortcut described in footnote \ref{footnote:weylrescale}) is that the Killing spinors $\xi_I, \tilde \xi_I$ describing supersymmetry variations should solve the Killing spinor equations\footnote{We follow the spinor conventions of Wess and Bagger \cite{wess1992supersymmetry}. See, for example, also appendix A of \cite{Pan:2015hza} for some more details. }
\begin{equation}\label{Killing-spinor-equations}
D_\mu \xi_I = - i \sigma_\mu \tilde \xi'_I\;, \qquad D_\mu \tilde \xi_I = - i \tilde\sigma_\mu \xi'_I \;,
\end{equation}
and the auxiliary equations
\begin{equation}\label{auxiliary-equations}
\sigma^\mu D_\mu \tilde \xi'_I = \frac{i}{4}M \xi _I\;,\qquad \sigma^\mu D_\mu \xi '_I = \frac{i}{4}M\tilde \xi_I\;.  
\end{equation}
The index $I$ carried by the Killing spinor is an $SU(2)_R$ index, and $M$ is a background field frozen to $M=-\frac{1}{3}\mathcal R$ with $\mathcal R$ the Ricci scalar, \ie, $\mathcal R=\frac{12}{r^2}$.\footnote{Note that for the case at hand the auxiliary equations are in fact a direct consequence of the Weitzenb\"ock formula.}  Let $\mathcal Q$ denote a supercharge associated to a Killing spinor solving \eqref{Killing-spinor-equations} and \eqref{auxiliary-equations}. The variations of a vector multiplet $(A_\mu, \phi, \tilde \phi, \lambda_I, \tilde \lambda_I, D_{IJ})$ are given by \cite{Hama:2012bg}
\begin{equation}\label{VM-SUSY}
\begin{aligned}
  \mathcal{Q}{A_\mu } = &\; + i({\xi ^I}{\sigma _\mu }{{\tilde \lambda }_I}) - i({{\tilde \xi }^I}{{\tilde \sigma }_\mu }{\lambda _I})  \\
  Q\phi  = &\;  - i({\xi ^I}{\lambda _I})  \\
  Q\tilde \phi  = &\;  + i({{\tilde \xi }^I}{{\tilde \lambda }_I})  \\
  \mathcal{Q}{\lambda _I} = &\; + \frac{1}{2}{F_{\mu \nu }}{\sigma ^{\mu \nu }}{\xi _I} + 2{D_\mu }\phi {\sigma ^\mu }{{\tilde \xi }_I} + \phi {\sigma ^\mu }{D_\mu }{{\tilde \xi }_I} + 2i{\xi _I}[\phi ,\tilde \phi ] + {D_{IJ}}{\xi ^J} \\
  \mathcal{Q}{{\tilde \lambda }_I} = &\; + \frac{1}{2}{F_{\mu \nu }}{{\tilde \sigma }^{\mu \nu }}{{\tilde \xi }_I} + 2{D_\mu }\tilde \phi {{\tilde \sigma }^\mu }{\xi _I} + \tilde \phi {{\tilde \sigma }^\mu }{D_\mu }{\xi _I} - 2i{{\tilde \xi }_I}[\phi ,\tilde \phi ] + {D_{IJ}}{{\tilde \xi }^J} \\
  \mathcal{Q}{D_{IJ}} = &\;  - i({{\tilde \xi }_I}{{\tilde \sigma }^\mu }{D_\mu }{\lambda _J}) + i({\xi _I}{\sigma ^\mu }{D_\mu }{{\tilde \lambda }_J}) - 2[\phi ,({{\tilde \xi }_I}{\lambda _J})] + 2[\tilde \phi ,({\xi _I}{\lambda _J})] + (I \leftrightarrow J)\;,
  \end{aligned}
\end{equation} 
while those of $N_f$ hypermultiplets $(q_{IA}, \psi_A, \tilde \psi_A, F_{IA})$ coupled to an $SU(N_f)$ vector multiplet $(A_\mu, \phi, \tilde \phi, \lambda_I, \tilde \lambda, D_{IJ})$ read
\begin{equation}\label{SUSY-variation}
\begin{aligned}
  \mathcal{Q}{q_{IA}} = &\; - i({\xi _I}{\psi _A}) + i({{\tilde \xi }_I}{{\tilde \psi }_A}) \\
  \mathcal{Q}{\psi _A} = &\; + 2{D_\mu }q_A^I{\sigma ^\mu }{{\tilde \xi }_I} - 4iq_A^I{\xi '_I} - 4 i \xi_I \tilde \phi{_A}{^B}q{^I}{_B} + 2{{\check \xi }_I}F{^I}_A \\
  \mathcal{Q}{{\tilde \psi }_A} = &\; + 2{D_\mu }q_A^I{{\tilde \sigma }^\mu }{\xi _I} - 4iq_A^I{\tilde \xi '_I} - 4 i \tilde \xi_I \phi{_A}{^B}q{^I}{_B}  + 2{\tilde {\check \xi} _I}F{^I}_A  \\
  \mathcal{Q}{F_{IA}} = &\; + i (\check \xi _I{\sigma ^\mu }{D_\mu }{{\tilde \psi }_A}) - i( \tilde{\check \xi}_I {{\tilde \sigma }^\mu }{D_\mu }{\psi _A} )\\
  & \; - 2 \phi{_A}{^B}(\check \xi_{I} \psi_B) - 2 (\check \xi_I \lambda_J){_A}{^B} q{^J}_B + 2 \tilde \phi{_A}{^B} (\tilde {\check \xi}_{I} \tilde \psi_B) + 2 (\tilde {\check \xi}_I \tilde\lambda_J){_A}{^B} q{^J}_B \;,
\end{aligned}
\end{equation}
Here $A,B,\ldots$ are $USp(2N_f)$ indices, and $SU(N_f)$ embeds in the standard way. The spinors ${\check \xi}_I, \tilde {\check \xi}_I$ should satisfy the constraints \cite{Hama:2012bg}
\begin{equation}\label{conditionsonauxspinors}
(\xi_I\check\xi_J)-(\tilde \xi_I\tilde{\check\xi}_J)=0\;, \quad (\xi^I\xi_I)+(\tilde {\check\xi}^I\tilde{\check\xi}_I)=0\;, \quad (\tilde\xi^I\tilde\xi_I)+( {\check\xi}^I{\check\xi}_I)=0\;, \quad (\xi^I \sigma^\mu \tilde\xi_I)+({\check\xi}^I\sigma^\mu\tilde{\check\xi}_I)=0\;.
\end{equation}
The supercharge $\mathcal Q$ squares to a sum of bosonic symmetries; see \cite{Hama:2012bg} for its detailed expression. In the next subsection, we will spell out $\mathcal Q^2$ for the Killing spinor of our interest.

The supersymmetric actions on the four-sphere read \cite{Pestun:2007rz,Hama:2012bg}
\begin{align}
&S_{\text{YM}}^{S^4} =\frac{1}{g_{\text{YM}}^2} \int d^4 x \sqrt{g_{S^4}} \Tr  \Big[ \frac{1}{2}F_{\mu\nu} F^{\mu\nu} - \frac{1}{2}D^{IJ}D_{IJ} -4D_\mu\tilde\phi D^\mu \phi -\frac{2\mathcal R}{3}\tilde \phi \phi + 4[\phi,\tilde\phi]^2  \nn\\
& \qquad\qquad\quad\qquad\qquad\qquad\qquad\qquad - 2i (\lambda^I\sigma^\mu D_\mu\tilde\lambda_I)-2(\lambda^I[\tilde\phi,\lambda_I]) + 2(\tilde\lambda^I[\phi,\tilde\lambda_I]) \Big]\;,\label{YMaction}\\
&{S_{{\text{HM}}}^{S^4}}=\int  {d^4}x \sqrt {g_{S^4}} \Bigg[ \frac{1}{2}{D_\mu }{q^{IA}}{D^\mu }{q_{IA}} - q^{IA} \{ \phi, \tilde \phi\}{_A}{^B} q_{IB} +\frac{i}{2}{q^{IA}}{D_{IJ}}{q^J}_A + \frac{\mathcal R}{12}{q^{IA}}{q_{IA}} \nn\\
& - \frac{i}{2}{{\tilde \psi }^A}{{\tilde \sigma }^\mu }{D_\mu }{\psi _A}  - \frac{1}{2}{\psi ^A}{\phi _A}^B{\psi _B} + \frac{1}{2}{\tilde \psi ^A}{\tilde \phi }{_A}^B{\tilde \psi _B}  - {q^{IA}}((\lambda _I)_A{^B}{\psi _B}) + ({\tilde \psi ^A}\tilde (\lambda ^I){_A}{^B}){q_{IB}}\Bigg]\;.  \label{HMaction}
\end{align}

\subsection{Choice of Killing spinor}\label{subsection:choiceofKS}
On the four-sphere, the equations \eqref{Killing-spinor-equations} (and \eqref{auxiliary-equations}) have sixteen $\mathbb C$-linearly independent solutions. These solutions are simply the Weyl transformations of the conformal Killing spinors on $\mathbb R^4$.\footnote{One should also perform a frame rotation to describe the spinors in the frame defined by the vielbeins \eqref{vielbeins}.} Among these spinors, we choose for our localization computation the Weyl transformation of the supercharge $\qq_1+\qq_2$. Concretely, the Killing spinor reads
\begin{align}
  & {\xi _{I = 1,\alpha }} =  \sin \frac{\rho }{2}\left( {\begin{array}{*{20}{c}}
    { - {e^{\frac{i}{2}( - \theta  + \varphi  + \chi )}}} \\ 
    {{e^{\frac{i}{2}(\theta  + \varphi  + \chi )}}} 
  \end{array}} \right)\;, & & {\xi _{I = 2,\alpha }} = \cos \frac{\rho }{2}\left( {\begin{array}{*{20}{c}}
    {{e^{\frac{i}{2}(\theta  + \varphi  - \chi )}}} \\ 
    {{e^{\frac{i}{2}( - \theta  + \varphi  - \chi )}}} 
  \end{array}} \right)\;,  \label{Killing-spinor-solutions1} \\
  & \tilde \xi _{I = 1}^{\dot \alpha } = i\sin \frac{\rho }{2}\left( {\begin{array}{*{20}{c}}
    {{e^{\frac{i}{2}(\theta  - \varphi  + \chi )}}} \\ 
    { - {e^{\frac{i}{2}( - \theta  - \varphi  + \chi )}}} 
  \end{array}} \right)\;, & & \tilde \xi _{I = 2}^{\dot \alpha } = i\cos \frac{\rho }{2}\left( {\begin{array}{*{20}{c}}
    {{e^{\frac{i}{2}( - \theta  - \varphi  - \chi )}}} \\ 
    {{e^{\frac{i}{2}(\theta  - \varphi  - \chi )}}} 
  \end{array}} \right)\;.
  \label{Killing-spinor-solutions2}
\end{align}
We can choose the auxiliary spinors $\check{\xi}_I$, $\tilde{\check \xi}_I$ as\footnote{The constraints \eqref{conditionsonauxspinors} do not uniquely fix the checked spinors. The ambiguity reflects the freedom to perform an independent $SU(2)_{\mathcal R'}$ rotation on the index of the checked spinors and the auxiliary field $F_{IA}$. Our choice identifies this $SU(2)_{\mathcal R'}$ and the standard $SU(2)_{\mathcal R}$. }
\begin{equation}
\check{\xi}_I = e^{-i \varphi} \xi_I\;,\qquad \tilde {\check{\xi}}_I = e^{+i \varphi} \tilde \xi_I\;.
\end{equation}

Let us define a few convenient bilinears built from the Killing spinors ${\xi}_I$, $\tilde{\xi}_I$
\begin{align}
& s \equiv ({\xi ^I}{\xi _I}) = -2 e^{i\varphi}\cos \theta \sin \rho,\qquad \tilde s \equiv ({{\tilde \xi }_I}{{\tilde \xi }^I}) = -2  e^{-i\varphi} \cos \theta \sin \rho,\\
& R^\mu \partial_\mu \equiv (\xi ^I\sigma ^\mu \tilde \xi _I) \partial_\mu = - \frac{2}{r}\partial_\varphi,\qquad R_{IJ}^\mu  \equiv (\xi _I\sigma ^\mu \tilde \xi _J)\;.
\end{align}

Let $\mathcal Q$ henceforth denote the supercharge described by the above Killing spinor. Its square can be computed straightforwardly:
\begin{equation}\label{Qsquared}
\mathcal Q^2 = -2i\mathcal L^{A}_R + \mathrm{Gauge}(2i(\tilde s\phi  + s\tilde \phi ))  + \mathcal R_{U(1)_r}(2r^{-1}) \;,
\end{equation}
where $\mathcal L^{A}_R$ is the gauge covariant Lie derivative along the vector field $R= - \frac{2}{r}\partial_\varphi$.


\section{Localizing the free hypermultiplet}\label{section:freeHM}
In this section, we localize the theory of a free hypermultiplet on the four-sphere with respect to the supercharge $\mathcal Q$ identified in the previous section. We will find that the theory localizes onto a two-dimensional quantum field theory on the two-sphere $S^2_{\theta=\frac{\pi}{2}}$. This quantum field theory precisely describes a symplectic boson pair, as expected from the algebraic discussion presented in section \ref{section:review-chiral-algebra}. The computation in this section is a crucial ingredient for our analysis in the next section of the effect of including defects. It also serves as a blueprint for the localization computation of gauge theories, which, however, is not the focus of the current paper and is left for future work.\footnote{See appendix \ref{app:VM} for a discussion of certain aspects of the extension to include vector multiplets.}

\subsection{Brief review of localization}\label{subsection:reviewLocalization}
Localization computations are based on the observation that in the path integral the action can be deformed by $\mathcal Q$-exact terms for any supercharge $\mathcal Q$ preserving the action, $\mathcal Q S=0$, \cite{Witten:1988ze,Witten:1991zz}\footnote{See also the recent comprehensive review \cite{Pestun:2016zxk}.}
\begin{equation}
\int [D\Phi] \ \mathcal O\  e^{-S[\Phi]}  = \int [D\Phi] \ \mathcal O \ e^{-S[\Phi] + t \mathcal Q \int V}\;, \qquad \text{if} \qquad \mathcal QS= \mathcal Q\mathcal O= \mathcal Q^2 \int V=0\;.
\end{equation}
Here $\Phi$ collectively denotes all fields in the theory, $V$ is any (local) fermionic functional of the quantum fields such that the integral of its variation under $\mathcal Q^2$ vanishes and $t$ is an arbitrary parameter. We also inserted an operator $\mathcal O$, local or non-local, order or disorder, which must be annihilated by $\mathcal Q$ for the equality to hold. Note that if the operator $\mathcal O$ can be obtained as the $\mathcal Q$-variation of another operator, the path integral will evaluate to zero, so the interesting observables lie in $\mathcal Q$-cohomology. If one now chooses $V$ such that the (bosonic part of) $\mathcal Q V$ is positive definite and sends $t$ to infinity, it is clear that the zeros of $\mathcal Q V$ will dominate the path integral. Let us denote this set of zeros as $\{\Phi_0\}$. The path integral localizes to this vanishing locus and it is sufficient to only take into account quadratic fluctuations of the deformation action $\mathcal Q \int V$ around them. In other words,
\begin{equation}
\int [D\Phi] \ \mathcal O \ e^{-S[\Phi]}  =\SumInt_{\left\{\Phi_0\right\}} \mathcal O|_{\Phi_0} \ e^{-S[\Phi_0]}\  Z_{\text{1-loop}}[\Phi_0]\;.
\end{equation}
A canonical choice for $V$ is\footnote{Other choices for $V$ have been analyzed in the literature in the context of so-called Higgs branch localization computations \cite{Benini:2012ui,Doroud:2012xw,Fujitsuka:2013fga,Benini:2013yva,Peelaers:2014ima,Pan:2014bwa,Chen:2015fta,Pan:2015hza}. It would be very interesting to study such other choices for the localizing supercharge used in the present paper.}
\begin{equation}
V_{\text{can.}} = \sum_{\text{fermions }\psi} (\mathcal Q \psi)^\dagger \psi\;,
\end{equation}
such that the localization locus $\{\Phi_0\}$ is given by the solutions of the BPS equations $\mathcal Q \psi =0$.

The localization computation thus consists of three steps:
\begin{enumerate}
\item find the localization locus, defined by the BPS equations $\mathcal Q \psi=0$
\item evaluate the classical action and any operator insertions on this locus
\item evaluate the one-loop determinant of quadratic fluctuations
\end{enumerate} 
We will address each of them in turn for the case of the free hypermultiplet.

\subsection{Localization locus}
The BPS equations can be read off from \eqref{SUSY-variation}, and we reproduce them here for convenience:
\begin{equation}
0 =  2{\partial_\mu }q_A^I{\sigma ^\mu }{{\tilde \xi }_I} - 4iq_A^I{\xi '_I}  + 2{{\check \xi }_I}F^I{_A}\;, \qquad 0 =  2{\partial_\mu }q_A^I{{\tilde \sigma }^\mu }{\xi _I} - 4iq_A^I{{\tilde \xi '}_I} + 2{\tilde {\check \xi} _I}F^I{_A}  \;.
\end{equation}
It is straightforward to deduce that
\begin{equation}
  \begin{aligned} 
    &\partial_\varphi q_{IA} =  \; 0 \;, \quad\qquad &{F_{JA}} &=   - \frac{1}{{\tilde s}}\Big((2{\check \xi _J}{\sigma ^\mu }{\tilde \xi _I}){D_\mu }{q^I}_A + ({\check \xi _J}{\sigma ^\mu }{D_\mu } \xi_I ){q^I}_A \Big)  \\
    && &=  + \frac{1}{s} \Big (2({\tilde {\check \xi} _J}{\tilde \sigma ^\mu }{\xi _I}){D_\mu }{q^I}_A + ({\tilde {\check \xi} _J}{\tilde \sigma ^\mu }{D_\mu }{\xi _I}){q^I}_A  \Big)\;.
  \end{aligned}
  \label{freeHM-BPS-equation-without-reality}
\end{equation}
Note that the first equation can also be read off from \eqref{Qsquared}, and that similarly $\partial_\varphi F_{IA}=0$.

Imposing the standard reality conditions $q_{IA}^\dagger = \epsilon^{IJ}\epsilon^{AB} q_{JB}$ and $F_{IA}^\dagger = - \epsilon^{IJ} \epsilon^{AB} F_{JB}$, the BPS equations can be split in their real and imaginary parts. We find explicitly for $A = 1$
\begin{align}
  0 = & \; \bigg[ \frac{+i}{\sin\theta\sin\rho}\partial_\chi q_{11} - \frac{\cos\theta}{\sin\rho}\partial_\theta q_{11} - \sin\theta\partial_\rho(\cos\rho q_{11}) \bigg] + i e^{i \chi} \partial_\rho(\sin\rho q_{21}) \nn\\
  0 = & \; \bigg[ \frac{-i}{\sin\theta\sin\rho}\partial_\chi q_{21} - \frac{\cos\theta}{\sin\rho}\partial_\theta q_{21} - \sin\theta\partial_\rho(\cos\rho q_{21}) \bigg] + i e^{ - i \chi} \partial_\rho(\sin\rho q_{11}) \nonumber\\
  F_{11} = &\; (r \cos\theta)^{-1} [- {e^{i\chi }}\sin \theta (r\sin \rho\; {\partial_4} + \cos \rho ){q_{21}} +(- r\sin \theta \; {\partial_2} - ir\cos \rho \; {\partial_4} + i\sin \rho ) {q_{11}}] \nn\\
  F_{21} = &\; (r \cos\theta)^{-1}[ - {e^{ - i\chi }}\sin \theta (r\sin \rho \; {\partial_4} + \cos \rho ){q_{11}} + ( r\sin \theta \; {\partial_2} - ir\cos \rho \; {\partial_4} + i\sin \rho){q_{21}}] \;,\label{BPS-equations}
\end{align}
while the equations for $A = 2$ can be obtained by complex conjugation of the above equations. It is clear that the first two equations constrain the hypermultiplet scalars, while the last two simply determine the auxiliary fields in terms of these scalars. We can thus focus on the first two equations. We will show momentarily that the moduli space of solutions to these equations (and their complex conjugates), and thus the space of BPS configurations, is one copy of the space of complex functions on $S^2_{\theta=\frac{\pi}{2}}$. We can thus already infer that the result of the localization computation will be a quantum field theory defined on $S^2_{\theta=\frac{\pi}{2}}$ describing the dynamics of this complex field. 

Let us now analyze the BPS equations for the hypermultiplet scalars $q_{11}$ and $q_{21}$, \ie{}, the first two equations of \eqref{BPS-equations}. Thanks to the $\varphi$-invariance of $q_{IA}$, they are coupled first-order partial differential equations on the three ball $D^3$ parametrized by $(\chi, \theta, \rho)$ and whose boundary is given by the sphere $S^2_{\theta = \frac{\pi}{2}}$. Let us first decompose the coupled system of equations into two independent second-order equations. To do so, we define the operators
\begin{align}
  \mathcal{D}_\pm \equiv \pm \frac{i}{\sin\theta\sin\rho }\partial_\chi - \frac{\cos\theta}{\sin\rho} \partial_\theta - \cos\rho \sin \theta \partial_\rho + \sin \rho \sin \theta\;,
\end{align}
and rewrite the equations \eqref{BPS-equations} as
\begin{equation}
0 =  e^{-i\chi}\mathcal{D}_+q_{11} + i \partial_\rho(\sin\rho q_{21})\;, \qquad 0 =  i e^{- i \chi} \partial_\rho(\sin\rho q_{11}) + \mathcal{D}_- q_{21}  \; .
\end{equation}
Let us then act on the left equation with the first-order differential operator $e^{+i\chi}(\mathcal{D}_- + \sin \rho \sin \theta)$, and on the right one with $ i e^{+i\chi} (\sin\rho\partial_\rho  + 2 \cos\rho)$. Subtracting the two resulting equations and using that
\begin{equation}
(\partial_\rho \sin\rho + \cos\rho)\mathcal{D}_\pm = (\mathcal{D}_\pm + \sin\rho \sin\theta) \partial_\rho \sin\rho \;,
\end{equation}
where the derivatives act on everything to the right, including any test function, one can extract a second-order, linear, homogeneous partial differential equation for $q_{11}$,
\begin{equation}
  0 = e^{+i\chi} (\mathcal{D}_ -  + \sin \rho \sin \theta )(e^{ - i\chi }\mathcal{D}_ + q_{11}) + (\partial _\rho \sin \rho  + \cos \rho )\partial _\rho (\sin \rho {q_{11}})\;.
\end{equation}
We recall that the qualitative behavior of the solutions to such a second-order linear differential equation is mostly controlled by its highest derivative terms:
\begin{equation}
  \bigg[\frac{1}{\sin^2\rho \sin^2\theta} \partial_\chi^2+ \frac{\cos^2 \theta}{\sin ^2\rho} \partial_\theta^2 + \frac{\cos\theta \cos\rho \sin\theta}{\sin\rho}(\partial_\theta \partial_\rho + \partial_\rho\partial_\theta) + (\sin^2 \rho + \sin^2\theta\cos^2\rho) \partial_\rho^2 \bigg]q_{11} \ . \nonumber
\end{equation}
Representing these terms as $A_{\bar \mu \bar \nu} \partial_{\bar \mu} \partial_{\bar \nu}$ with $\bar \mu = \chi, \theta, \rho$, it immediately follows that $\det A = \cos^2\theta (\sin\theta\sin\rho)^{-2} \ge 0$. This is precisely the ellipticity condition of the second-order differential equation for $q_{11}$. Similar to the Dirichlet problem of the Laplace equation, the solutions to the equation for $q_{11}$ are thus determined by the value of $q_{11}$ at the boundary sphere $S^2_{\theta = \frac{\pi}{2}}$ (see for example \cite{Gilbarg}). An almost identical analysis can be performed for $q_{12}$. Note however, that the original equations should still hold, restricting the freedom in the solution to $q_{12}$. This shows that the moduli space of the BPS equations (\ref{BPS-equations}) is indeed given by the space of complex functions on $S^2_{\theta = \frac{\pi}{2}}$.

\subsection{Evaluation of the classical action}\label{subsection:classical-action-freeHM}
Our next task is the evaluation of the classical action \eqref{HMaction}\footnote{In this section we are considering a free hypermultiplet and hence we consider the action in the absence of its coupling to the vector multiplet. In appendix \ref{app:evaluateCA} we consider the evaluation of both the Yang-Mills action and the gauged hypermultiplet action on configurations satisfying the (complex) BPS equations.} on the BPS configurations. First of all, we argued above that \eqref{Qsquared} implies that $q_{IA}, F_{IA}$ are independent of the coordinate $\varphi$. Hence, the (bosonic part of the) free hypermultiplet Lagrangian is independent of the coordinate $\varphi$, and we can perform the $\varphi$-integral effortlessly. As a result, we are left with an integral over a three-ball $D^3$, parametrized by the coordinates $\chi, \theta, \rho$, with boundary $\partial D^3 = S^2_{\theta = \frac{\pi}{2}}$.\footnote{The coordinate $\varphi$ is ill-defined on this locus. We set it to zero consistently.}  Next we substitute the complex BPS solutions for the auxiliary fields $F_{IA}$ as in equation \eqref{BPS-equations}.\footnote{As was observed in for example \cite{Dedushenko:2016jxl,Bonetti:2016nma}, it is sufficient to consider the complex BPS equations to evaluate the classical action and reduce it to an action of a lower-dimensional quantum field theory.} It is straightforward to show that the hypermultiplet Lagrangian can then be written as a total divergence,
\begin{align}
  2\pi r \sqrt {{g_{D^3}}} {\mathcal{L}_{{\text{HM}}}} = & \; 2\pi r \sqrt {g_{D^3}} {\nabla _{\bar \mu} }\Big[ {\frac{1}{{s\tilde s}}{\epsilon ^{{\bar \mu} \nu \lambda \delta }}{R_\lambda }{{({R_{IJ}})}_\delta }{q^I}_A{D_\nu }{q^{JA}} - \frac{{2i}}{{\tilde s}}(\xi '_I \sigma ^{\bar \mu} {{\tilde \xi }_J}){q^I}_A{q^{JA}}} \Big] \nn \\
  = & \; 2\pi r \sqrt{g_{D^3}}\nabla_{\bar \mu}\Big[ \frac{1}{{s\tilde s}}{\epsilon ^{{\bar \mu} \nu \lambda \delta }}{R_\lambda }(({q^I}_A{\xi _I}){\sigma _\delta }{D_\nu }({q^{JA}}{{\tilde \xi }_J})) - \frac{{2i}}{{s\tilde s}}{R^{\bar \mu} }({\xi _I}{\xi '_J}){q^I}_A{q^{JA}} \Big] \nn\\
  \equiv & \; 2\pi r \sqrt{g_{D^3}}\nabla_{\bar \mu} K^{\bar \mu}\;, \qquad \qquad \bar \mu= \chi, \theta, \rho\;.
\end{align}
Here $\sqrt {{g_{{D^3}}}}  = {r^3}\cos \theta \sin \theta {\sin ^3}\rho $ is the square root of the determinant of the metric on the three-ball and ${\epsilon _{\mu \nu \lambda \delta }} \equiv e_\mu ^ae_\nu ^be_\lambda ^ce_\delta ^d\epsilon _{abcd}$ is the standard volume form. Using Stokes' theorem, one can reduce the integral over the three-ball to an integral over the boundary. The latter can be reinterpreted as the action for the lower-dimensional quantum field theory onto which we are localizing. In other words,
\begin{equation}
S^\text{2d}_{{\text{HM}}} = 2\pi r \int {d\rho d\chi \sqrt {{g_{{S^2}}}}\; \mathcal{L}_{{\text{HM}}}^{{\text{2d}}}} \;,\qquad \text{with}\qquad \mathcal{L}_{{\text{HM}}}^{{\text{2d}}} = {[ {{{(r^2\sin\rho )}^{ - 1}}\sqrt {{g_{{D^3}}}} {K^\theta }} ]_{\theta  \to \pi /2}}\;.
\end{equation}

To write the Lagrangian $\mathcal{L}_{{\text{HM}}}^{{\text{2d}}}$ more concretely, it is convenient to first define two flavor doublets of spinors $\mathfrak{q}_A\equiv q{^I}{_A} \xi_I$, $\tilde {\mathfrak{q}}_A \equiv q{^I}_A \tilde \xi_I$. As explained in detail in appendix \ref{subsection:reducespinorsontothetapiotwo}, when restricting to $S^2_{\theta=\frac{\pi}{2}}$, these spinors can be traded for a two-dimensional anti-chiral spinor $\mathfrak{q}_A^\text{2d}$.\footnote{When restricting a pair of four-dimensional spinors $\psi_\alpha, \tilde \psi_{\dot\alpha}$ to a two-dimensional locus, one typically finds a pair of two-dimensional spinors $\psi^{\text{2d}},\tilde\psi^{\text{2d}}$. It turns out that for the the case of $\mathfrak{q}_{\alpha A},\tilde {\mathfrak{q}}_{\dot\alpha A}$ the two-dimensional spinors $\mathfrak{q}_A^\text{2d}$ and $\tilde{\mathfrak{q}}_A^\text{2d}$ are proportional to each other and anti-chiral in the two-dimensional sense.} Concretely, in the patch $U_{\text{eq}}$ of the two-sphere excluding both the north pole and south pole, the spinor $\mathfrak{q}_A^\text{2d}$ reads\footnote{\label{footnotePatches}We complete the patch $U_{\text{eq}}$ to an open cover of the two-sphere as $S^2 = U_\text{N} \cup U_\text{S} \cup U_\text{eq}$, where $U_{\text{N}/\text{S}}$ excludes the south/north pole of the two-sphere. We choose the standard vielbein \eqref{vielbeinsS2thetapi2} on $U_\text{eq}$. The vielbein on $U_\text{N/S}$ are related by a frame rotation
\begin{align}
  (e^a)_\text{N} \equiv (U^{-1}){^a}{_b}\; (e^b)_\text{eq}\;, \quad (e^a)_\text{S} \equiv U{^a}{_b}\; (e^b)_\text{eq}\;, \quad \text{with }\quad U =  
  \left(
    \begin{array}{*{20}{c}}
      \cos \chi &\sin \chi  \\ 
       - \sin \chi &\cos \chi  
    \end{array}
  \right)\;.
\end{align}
and are simply induced from stereographic projection. The anti-chiral component $Q_A$ of the spinor thus satisfies
\begin{equation}\label{changepatchonQA}
Q_A^{(\text{N})} = e^{\frac{i\chi}{2}}\; Q_A^{(\text{eq})}\;,\qquad Q_A^{(\text{S})} = e^{-\frac{i\chi}{2}}\; Q_A^{(\text{eq})}\;.
\end{equation}}
\begin{equation}\label{definitionQA}
\mathfrak{q}_A^\text{2d,(eq)} = \begin{pmatrix}
0 \\
Q_A^{\text{(eq)}}
\end{pmatrix} = \begin{pmatrix}
0 \\
e^{-\frac{i\chi}{2}} \cos \frac{\rho}{2} \ q_{1A} - i e^{\frac{i\chi}{2}}\sin \frac{\rho}{2}\ q_{2A}
\end{pmatrix}\;.
\end{equation}
The fully covariant two-dimensional Lagrangian is
\begin{equation}\label{2dlagrangianforHM}
\mathcal L_\text{HM}^\text{2d}[\mathfrak{q}] = - \epsilon^{AB} {\epsilon ^{\mu\nu}}({\mathfrak{q}^\text{2d}_A}{\gamma _\mu}{\nabla^\text{2d}_\nu}\mathfrak{q}^\text{2d}_B)\;, \qquad \mu, \nu = \rho, \chi\;.
\end{equation}
Here $\gamma_a$ are two-dimensional gamma-matrices,\footnote{\label{footnote2dgamma}We take $\gamma_1 = \tau_1$, $\gamma_2=\tau_2$, and thus $\gamma_3=\tau_3$, in terms of the Pauli-matrices and define the inner product between two two-dimensional spinors as $(\psi\chi)\equiv \psi^T (i\tau_2) \chi$.} while $\nabla^{\text{2d}}$ denotes the two-dimensional covariant derivative on the two-sphere, \ie{}, $\nabla^\text{2d}_\nu\mathfrak{q}^\text{2d}_B = \partial_\nu\mathfrak{q}^\text{2d}_B + \frac{1}{4}\omega_{\nu}^{\phantom{\nu}ab}\gamma_{ab}\mathfrak{q}^\text{2d}_B$, with $\omega_{\nu}^{\phantom{\nu}ab}$ the standard spin-connection on the two-sphere.

In spinor components, and defining $Q^{(\text{eq})}\equiv Q_1^{(\text{eq})}$ and $\widetilde Q^{(\text{eq})} \equiv Q_2^{(\text{eq})}$, the action can be written alternatively as
\begin{equation}
  S^\text{2d}_{{\text{HM}}} = - 4\pi i\int d\chi d\rho \sqrt {g_{S^2}}\ \widetilde Q^{(\text{eq})}\; \overline{\wp}^{(\text{eq})} \; Q^{(\text{eq})}\;,
\end{equation}
where $\overline{\wp}^{(\text{eq})}$ is the lowering operator for the magnetic charge $\mathfrak{m}=\frac{1}{2}$ of monopole harmonics in the equator patch \cite{Wu:1976ge,Dray:1984gy}. Its expression, as well as those in the patches $U_\text{N/S}$ which exclude the south/north pole respectively (see also footnote \ref{footnotePatches}), is
\begin{equation}
\overline{\wp}^{(\text{eq})} \equiv \Big(\partial_\rho - \frac{i}{\sin\rho}\partial_\chi + \mathfrak{m}\frac{\cos\rho}{\sin\rho} \Big)\;, \qquad \overline{\wp}^{(\text{N/S})} \equiv e^{\mp i\chi}\Big(\partial_\rho - \frac{i}{\sin\rho}\partial_\chi + \mathfrak{m}\frac{\cos\rho \mp 1}{\sin\rho} \Big)\;.
\end{equation}
With these operators, the action can be written uniformly in all patches:
\begin{equation}\label{2dlagrangianforHMinComp}
S^\text{2d}_{{\text{HM}}} = - 4\pi i\int d\chi d\rho \sqrt {g_{S^2}}\ \widetilde Q\; \overline{\wp} \; Q\;.
\end{equation}
Note that the fields $Q$ and $\widetilde Q$ are related by a reality condition. This follows directly from the analysis in the previous subsection.

\subsection{Evaluation of the one-loop determinant}\label{subsec: evaluation oneloop}
The final step in the localization computation is to evaluate the one-loop determinant of quadratic fluctuations. We will argue that
\begin{equation}
Z_{\text{1-loop}} = 1\;.
\end{equation}
We first observe that the deformation action is quadratic in the hypermultiplet fields. The differential operator describing quadratic fluctuations around the BPS configurations, and therefore also the one-loop determinant, can thus only depend on background parameters.\footnote{Here we make use of the fact that the hypermultiplet is free. For the interacting case, the one-loop determinant can depend nontrivially on the localization locus of the vector multiplet.} Consequentially, the one-loop determinant provides an overall normalization to the partition function which, while immaterial for the computation of correlation functions, can be fixed by demanding that the partition function of the two-dimensional theory defined by the action \eqref{2dlagrangianforHM} or \eqref{2dlagrangianforHMinComp} equals the partition function of the four-dimensional free hypermultiplet. Both computations are straightforward, as the theories are Gaussian. In fact, the result of the latter can be borrowed from the localization literature \cite{Pestun:2007rz,Hama:2012bg} and reads
\begin{equation}
  {Z^{{\text{HM}}}_{\text{1-loop}}} =\lim_{\substack{x \to 0 \\ b \to 1}} \Upsilon_b\Big(\frac{b+b^{-1}}{2} + x\Big)\;,
\end{equation}
in terms of Upsilon-function $\Upsilon_b$.\footnote{The Upsilon-function is the regularization of the infinite product
\begin{equation}
\Upsilon_b(x) = \prod_{m,n\geq 0} (mb+nb^{-1}+x) ((m+1)b+(n+1)b^{-1}-x) \;.
\end{equation}
}

Now let us compute the partition function of the two-dimensional theory. It is natural to decompose $Q$ and $\widetilde Q$ in monopole harmonics $Y_{jm}^{\mathfrak{m}=1/2}$. Indeed, given the monopole charge $\mathfrak{m}=\frac{1}{2}$, $\{Y^\mathfrak{m}_{jm} | j \in \mathbb{N} + |\mathfrak{m}|, m \in \{ - j, -j + 1, \ldots, +j\} \}$ forms an orthonormal basis for the space of sections $\Gamma(\mathcal{S}_-)$ of the nontrivial anti-chiral spinor bundle $\mathcal{S}_-$.\footnote{On $S^2$ there are two inequivalent spin structures. Here the spinor bundle corresponds to the nontrivial spin structure.} The first order differential operator $\overline{\wp}$ can be viewed as $\overline{\partial} - i a_{\bar z}$ on $\Gamma(\mathcal{S}_-)$, where $a$ is the $U(1)$ connection on $\mathcal S_-$, taking the monopole profile $a = \frac{1}{2}(\pm1-\cos\rho) d\chi$ on $U_\text{N/S}$. It is easy to work out the eigenvalue of the operator $\overline{\wp}$ by working in a patch. One finds
\begin{equation}
\overline{\wp}  Y_{jm}^{\mathfrak{m}} = - \sqrt {(j + \mathfrak{m})(j + 1 - \mathfrak{m})} Y_{jm}^{\mathfrak{m}-1}\quad \Longrightarrow\quad  \overline{\wp}  Y_{jm}^{\mathfrak{m} = 1/2} = - (j + \frac{1}{2}) Y_{jm}^{\mathfrak{m} = -1/2}\;.
\end{equation}
Note that $\overline{Y^\mathfrak{m}_{jm} (\rho, \chi)} = (-1)^{\mathfrak{m} + m} \, Y^\mathfrak{ - m}_{j, -m} (\rho, \chi)$ and thus $\overline{Y^\mathfrak{m}_{jm}}$ and $Y^\mathfrak{ - m}_{j, -m}$ are sections of the same line bundle $\mathcal{S}_- $. We take
\begin{equation}
Q = \sum_{j \in \mathbb{N} + 1/2} {\sum_{m =  - j}^{ + j} {Q_{jm}\  Y_{jm}^{\mathfrak{m} = 1/2}} }\;,\qquad \widetilde Q = \sum_{j \in \mathbb{N} + 1/2} {\sum_{m =  - j}^{ + j} {\widetilde Q_{jm}\  \overline {Y_{jm}^{\mathfrak{m} = - 1/2}} } } \;,
\label{expanding-Q}
\end{equation}
and impose $\overline{Q_{jm}} = \widetilde Q_{jm}$.\footnote{Recall that indeed $Q$ and $\widetilde Q$ are related by a reality property.} Using the orthogonality condition
\begin{equation}
\int {d\rho d\chi \sqrt {{g_{{S^2}}}}\ \overline {Y_{jm}^{ - 1/2}} \; Y_{j'm'}^{ - 1/2}}  = {\delta _{jj'}}{\delta _{mm'}}\;,
\end{equation}
the two-dimensional action simplifies to
\begin{equation}
S^\text{2d}_{{\text{HM}}} = 4\pi i\sum\limits_{j = 1/2}^{ + \infty  + 1/2} {\sum\limits_{m =  - j}^{ + j} {(j + 1/2){Q_{jm}}\overline {{Q_{jm}}} } } \;.
\end{equation}
The partition function can now be computed easily and reads, up to some normalization constants,
\begin{align}
Z^\text{2d}_{{\text{HM}}}  = & \; \prod\limits_{j = 0}^{ + \infty } {{{(j + 1)}^{2j + 2}}}  = \mathop {\lim } _{\substack{x \to 0 \\ b \to 1}} \prod\limits_{m,n = 0}^{ + \infty } {(mb + n{b^{ - 1}} + Q/2 + x)(mb + n{b^{ - 1}} + Q/2 - x)}\nn \\
  = & \; \mathop {\lim } _{\substack{x \to 0 \\ b \to 1}} \Upsilon\Big(\frac{Q}{2} + x\Big)\;.
\end{align}
This shows that the one-loop determinant is indeed equal to one: $Z_\text{1-loop}= 1$.

\subsection{Final result}\label{subsection:finalResultHMlocalization}
The analysis of the previous few subsections shows that the free hypermultiplet on the four-sphere can be localized to a two-dimensional quantum field theory, described by the action $S^\text{2d}_{{\text{HM}}}$ in equation \eqref{2dlagrangianforHMinComp}. In formulae,
\begin{equation}
\int [D\Phi_{\text{HM}}] \ \mathcal O\  e^{-S_{\text{HM}}[\Phi_{\text{HM}}]} = \int [DQ D\widetilde Q] \ \mathcal O|_{Q,\widetilde Q} \ e^{-S^\text{2d}_{{\text{HM}}}[Q,\widetilde Q]}\;, \qquad \text{if} \qquad \mathcal Q\mathcal O= 0\;,
\end{equation}
where $\Phi_{\text{HM}}$ collectively denotes all fields in the hypermultiplet. The localization argument only guarantees the above equality if the observables $\mathcal O$ are annihilated by $\mathcal Q$. As the two-dimensional theory describes the dynamics of $Q$ and $\widetilde Q$, it is natural to expect that, as far as local operators go, composites of these fields are good observables on the two-sphere. Indeed, it is easy to verify that 
\begin{equation}
\mathcal Q Q \big|_{S^2_{\theta=\frac{\pi}{2}}} = 0\;, \qquad \mathcal Q \widetilde Q \big|_{S^2_{\theta=\frac{\pi}{2}}} = 0\;,
\end{equation}
and similarly for any composite made from the letters $Q$, $\widetilde Q$ and the operator $\wp$.\footnote{Note that if one wants to relate correlation functions computed on the four-sphere to those on flat space, one should construct a suitable basis of composite primary operators, see \cite{Gerchkovitz:2016gxx}.} It may be useful to note that the combinations $Q$ and $\widetilde Q$ of elementary four-dimensional fields, see \eqref{definitionQA}, can also be obtained by performing a $\mathcal Q$-closed (twisted) translation as on flat space (see section \ref{section:review-chiral-algebra}).

The two-dimensional theory is Gaussian, hence all correlators can be computed via Wick contractions. The key-ingredient is thus the propagator $G(\rho ,\chi ;\rho ',\chi ') =\langle Q(\rho ,\chi)\ \widetilde Q(\rho ',\chi ')\rangle = - \langle \widetilde Q(\rho ,\chi)\ Q(\rho ',\chi ')\rangle$. It should solve the differential equation
\begin{equation}
\overline{\wp} \; G(\rho ,\chi ;\rho ',\chi ') =  -\frac{1}{4\pi i} {\delta_{{S^2}}}(\rho  - \rho ',\chi  - \chi ')\;,
\end{equation}
where ${\delta_{{S^2}}}(\rho - \rho' ,\chi - \chi')$ is the Dirac $\delta$-function on $S^2$, \ie{},
\begin{equation}
\int_{S^2} d\rho d\chi \sqrt {{g_{{S^2}}}}\ f(\rho,\chi) {\delta_{{S^2}}}(\rho - \rho' ,\chi - \chi') = f(\rho',\chi')\;.
\end{equation}
It is easy to show that in the equator patch
\begin{align}\label{propagator}
G^{(\text{eq})}(\rho, \chi; \rho', \chi') = \frac{-2i}{(4\pi r)^2}\; \frac{{ {{e^{-\frac{i}{2}(\chi-\chi')}}\cos \left( {\frac{\rho }{2}} \right)\sin \left( {\frac{{\rho '}}{2}} \right) - {e^{\frac{i}{2}(\chi-\chi') }}\sin \left( {\frac{\rho }{2}} \right)\cos \left( {\frac{{\rho '}}{2}} \right)} }}{{1 - (\sin \rho \sin \rho '\cos (\chi - \chi ')+\cos \rho \cos \rho ')}}
\end{align}
is the solution.

There are two shortcuts to construct the above propagator. The first one leverages the Weyl transformation to flat space, discussed around \eqref{WeylRescaling}. Starting form the standard flat space propagator of a symplectic boson pair, multiplying the appropriate conformal factors, performing a change of coordinates and finally rotating the frame as in \eqref{changepatchonQA} results in \eqref{propagator}. The second one starts from the propagator of a massive free scalar field on the four-sphere. If one sets the mass to its conformal value, considers the combinations $Q_A^{\text{(eq)}}$ as in \eqref{definitionQA} and restricts oneself to the two-sphere $S^2_{\theta=\frac{\pi}{2}}$, the result \eqref{propagator} also follows. Let us give a few more details on the latter computation. The propagator for a free real scalar field of mass $m$ on a four-sphere of radius $r$ solves
\begin{equation}\label{S4scalarpropagator}
(-\nabla^\mu \partial_\mu + m^2)\ G_{S^4}(x,x';m^2) = \delta_{S^4}(x-x')\;,
\end{equation}
where the delta-function is with respect to the four-sphere measure. The solution to \eqref{S4scalarpropagator} is given by \cite{Allen:1985wd}
\begin{equation}
G_{S^4}(x,x';m^2) = \frac{1}{(4 \pi r)^2} \Gamma(3/2+\nu)\Gamma(3/2-\nu)\  {}_2F_1\left(3/2+\nu,3/2-\nu; 2; \cos^2\left(\frac{\mu(x,x')}{2r}\right) \right)\;, 
\end{equation}
where $\nu$ is computed by $\nu^2 = \frac{9}{4} - (mr)^2$, and $\mu(x,x')$ denotes the geodesic distance between the points $x$ and $x'$. This latter quantity is easily computed to be
\begin{small}
\begin{equation}
\mu(x,x') =r \arccos\left(\cos\rho\,\cos\rho' + \sin\rho\ \sin\rho'\left(\cos\theta\,\cos\theta'\ \cos(\varphi-\varphi') + \sin\theta\, \sin\theta'\, \cos(\chi-\chi')  \right) \right)\;,
\end{equation}
\end{small}%
if $x=(\varphi,\chi,\theta,\rho)$ and $x'=(\varphi',\chi',\theta',\rho')$. As we are interested in conformally coupled scalars, we set
\begin{equation}
m^2 = \frac{\mathcal R}{6} = \frac{2}{r^2}\;, \qquad \nu = \frac{1}{2}\;.
\end{equation}
Noting that ${}_2F_1\left(2,1; 2; y \right) = \frac{1}{1-y}$ and $\cos^2\left(\frac{1}{2} \arccos t\right) = \frac{1+t}{2}$, the propagator simplifies significantly. Finally, restricting the insertion points to the two-sphere at $\theta=\pi/2$ and considering the combinations $Q, \widetilde Q$ as in \eqref{definitionQA}, one recovers \eqref{propagator}.

In section \ref{section:review-chiral-algebra}, we reviewed that the chiral algebra associated to a free hypermultiplet is a symplectic boson pair. Let us show that the action $S^\text{2d}_{{\text{HM}}} = \int d^2x \sqrt{g_{S^2}}\ \mathcal L^\text{2d}_{{\text{HM}}}$ indeed describes a curved space version of the symplectic boson. We do so by performing a Weyl transformation. It is straightforward to verify that the Lagrangian density \eqref{2dlagrangianforHM} multiplied by the measure $\sqrt{g_{S^2}}$ is invariant under the Weyl rescalings
\begin{equation}\label{WeylRescaling}
g_{\mu\nu}\rightarrow \Omega^2 g_{\mu\nu}\;, \qquad \mathfrak q_A \rightarrow \Omega^{-1/2} \mathfrak q_A\;.
\end{equation}
Let us then perform a Weyl recaling from $S^2$ to $\mathbb R^2=\mathbb C$. In terms of the component $Q_A$ of $\mathfrak q_A$, and using that $d^2 x = -\frac{i}{2} dz d\bar z$, the action becomes
\begin{equation}
S^{\text{2d,}\mathbb C}_{{\text{HM}}} = \frac{1}{2}\int dzd\bar z\ \epsilon^{AB} Q_A \,\partial_{\bar z} Q_B = -\int dzd\bar z\ \widetilde Q\, \partial_{\bar z} Q\;,
\end{equation}
where once again $Q=Q_{A = 1}$ and $\widetilde Q = Q_{A = 2}$. One can readily recognize this action as describing a symplectic boson pair. 

The symplectic boson is a special instance of the more general $\beta\gamma$-system, where $\beta$ and $\gamma$ have equal conformal weight. The $\beta\gamma$-system can be placed canonically on an arbitrary Riemann surface $\Sigma$
\begin{equation}\label{betagammaaction}
S_{\beta\gamma} = \int_{\Sigma} \beta \bar\partial \gamma\;,  \qquad \text{with}\qquad \beta \in \Omega^{1,0}(\Sigma), \quad \gamma \in \Omega^{0,0} (\Sigma) \;.
\end{equation}
As we now argue, our action \eqref{2dlagrangianforHMinComp} on the two-sphere is related to this action via a topological twist. The action \eqref{2dlagrangianforHMinComp} is written in terms of anti-chiral spinor components and it possesses a $U(1)$ flavor symmetry.\footnote{More precisely, it has an $SU(2)$ flavor symmetry. We focus on its Cartan here.} Denoting the anti-chiral spinor bundle as $S_-$ and the flavor line bundle as $L$, the field $Q$ is a section of $S_-\otimes L$ and $\widetilde Q$ is a section of $S_-\otimes L^{-1}$. Performing a topological twist, identifying $L^{-1}$ with the line bundle $S_-$, and using that $S_- = (\bigwedge^{1,0})^\frac{1}{2}$, one can turn $Q$ into a section of $\bigwedge^{0,0}$ and $\widetilde Q$ into one of $\bigwedge^{1,0}$. The resulting action is then precisely \eqref{betagammaaction}.


\section{Surface defects and chiral algebra modules}\label{section:defects}
In the previous section, we have localized the theory of a free hypermultiplet to a two-dimensional quantum field theory on $S^2_{\theta = \frac{\pi}{2}}$. The computation is compatible with the insertion of operators preserving the localizing supercharge. We have already considered the insertion of local operators, and will now turn attention to non-local operators. We consider the insertion of a surface defect preserving two-dimensional $\mathcal N=(2,2)$ supersymmetry on the two-sphere $S^2_{\theta = 0}$. We describe the defect by coupling the four-dimensional theory to additional degrees of freedom residing on $S^2_{\theta = 0}$.

\subsection{Embedding 2d \texorpdfstring{$\mathcal N=(2,2)$}{N=(2,2)} into 4d \texorpdfstring{$\mathcal N=2$}{N=2}}\label{embeddingflatspace}
To couple lower-dimensional degrees of freedom to a higher-dimensional quantum field theory, it is convenient to decompose the higher-dimensional multiplets in terms of lower-dimensional ones while treating the extra coordinates as continuous labels. Indeed, such decomposition trivializes the task of writing the action coupling the two systems while manifestly preserving the lower-dimensional symmetries, in particular supersymmetry. Naturally, the decomposition depends on the embedding of the lower-dimensional algebra into the higher-dimensional one. Our first order of business is thus to specify the relevant embedding of the two-dimensional $\mathcal N=(2,2)$ superconformal algebra in the four-dimensional $\mathcal N=2$ superconformal algebra. For convenience, we do so in flat space; a Weyl transformation easily maps the result to the four-sphere.

We start by briefly describing the relevant superalgebras. The non-zero anticommutation relations among the fermionic generators $\{\mathcal Q_{I\alpha}, \widetilde Q^I_{\dot\alpha}, \mathcal S^{I\alpha}, \widetilde{\mathcal S}_I^{\dot \alpha} \}$ of the four-dimensional $\mathcal N=2$ superconformal algebra, \ie{}, $\mathfrak{su}(2,2|2)$, are given by
\begin{align}
&\{\mathcal Q_{I\alpha}, \widetilde {\mathcal Q}^J_{\dot\alpha}\} = \delta^J_I P_{\alpha\dot\alpha}\;, \qquad \{\widetilde{\mathcal S}_{I}^{\dot\alpha}, {\mathcal S}^{J\alpha}\} = \delta^J_I K^{\dot\alpha\alpha}\;, \\
&\{\mathcal Q_{I\alpha},\mathcal S^{J\beta} \}=\frac{1}{2}\delta^J_I \delta_\alpha^\beta \Delta + \delta^J_I \mathcal{M}_\alpha^{\phantom{\alpha}\beta}-\delta_\alpha^\beta \mathcal R_I^{\phantom{I}J}\;, \\
&\{\widetilde{\mathcal S}_{I}^{\dot\alpha},\widetilde{\mathcal Q}^{J}_{\dot\beta} \}=\frac{1}{2}\delta^J_I \delta^{\dot\alpha}_{\dot\beta} \Delta + \delta^J_I \mathcal{M}^{\dot\alpha}_{\phantom{\alpha}\dot\beta}+\delta^{\dot\alpha}_{\dot\beta} \mathcal R_I^{\phantom{I}J}\;.
\end{align}
Here $P_{\alpha\dot\alpha}$ and $K^{\dot\alpha\alpha}$ are the usual generators for translation and special conformal transformation, $\Delta$ denotes the dilatation generator, and $\mathcal{M}_\alpha^{\phantom{\alpha}\beta}$ and $\mathcal{M}^{\dot\alpha}_{\phantom{\alpha}\dot\beta}$ are the rotational generators of $\mathfrak{su}(2)_1 \oplus \mathfrak{su}(2)_2=\mathfrak{so}(4)$. Together, $P_{\alpha\dot\alpha},K^{\dot\alpha\alpha},\Delta,\mathcal{M}_\alpha^{\phantom{\alpha}\beta}$ and $\mathcal{M}^{\dot\alpha}_{\phantom{\alpha}\dot\beta}$ generate the four-dimensional conformal algebra $\mathfrak{su}(2,2)$. Finally, $\mathcal R_I^{\phantom{I}J}$ generate the R-symmetry $\mathfrak{su}(2)_R \oplus \mathfrak{u}(1)_r$. In particular, one has
\begin{equation}
\mathcal M^\bot = \mathcal M_+^{\phantom{+}+}-\mathcal M^{\dot +}_{\phantom{+}\dot +}\;, \qquad \mathcal M = \mathcal M_+^{\phantom{+}+}+\mathcal M^{\dot +}_{\phantom{+}\dot +}\;, \qquad \mathcal R_1^{\phantom{1}1}=\frac{1}{4}r + R\;, \qquad \mathcal R_2^{\phantom{2}2}=\frac{1}{4}r - R\;,
\end{equation}
expressing the generators rotating the $(x_1,x_2)$-plane and $(x_3,x_4)$-plane respectively, and relating the diagonal R-symmetry generators to the $SU(2)_R$ Cartan generator $R$ and the $U(1)_r$ charge $r$.

The two-dimensional $\mathcal N=(2,2)$ superconformal algebra is $\mathfrak{su}(1,1|1)_L\oplus \mathfrak{su}(1,1|1)_R$. The bosonic subgroup of one copy of the algebra, say $\mathfrak{su}(1,1|1)_L$, is $\mathfrak{su}(1,1)_L\oplus \mathfrak{u}(1)_L$. Here $\mathfrak{su}(1,1)_L$ is the spatial part, with standard generators $L_0, L_{\pm 1}$, and $\mathfrak{u}(1)_L$ is the left-moving R-symmetry generated by $J_0$. We denote the left-moving Poincar\'e supercharges as $G_{- \frac{1}{2}}^\pm$ and conformal supercharges as $G_{+ \frac{1}{2}}^\pm$. Some of the (anti)commutation relations involving the supercharges are
\begin{equation}
[L_0,G_r^\pm] = -r G_r^\pm\;, \quad [J_0,G_r^\pm] = \pm G_r^\pm\;, \quad \{G_r^+,G_s^- \} = L_{r+s} + \frac{r-s}{2}J_{r+s}\;, \quad \text{for } r,s=\pm\frac{1}{2}\;. 
\end{equation}
We will denote the generators of the right-moving algebra $\mathfrak{su}(1,1|1)_R$ with a bar.

The two-dimensional $\mathcal N=(2,2)$ superconformal algebra can be embedded in the four-dimensional $\mathcal N=2$ superconformal algebra in various ways. Let us choose to orient the embedded plane along the $(x_1,x_2)$-directions. For the purposes of this paper, we make the following identifications between two-dimensional supercharges and four-dimensional supercharges:\footnote{This embedding has, for example, also been described in detail in appendix A of \cite{Cordova:2017ohl}.}
\begin{equation}
\begin{aligned}
&G^+_{-\frac{1}{2}} = \mathcal Q_{2+}\;, \qquad &&G^-_{-\frac{1}{2}} = \widetilde{\mathcal Q}^2_{\dot -}\;, \qquad &&\bar G^+_{-\frac{1}{2}}=\mathcal Q_{1-}\;, \qquad &&\bar G^-_{-\frac{1}{2}}=\widetilde{\mathcal Q}^1_{\dot +}\;, \\
&G^+_{+\frac{1}{2}} = \widetilde{\mathcal S}_{2}^{\dot -}\;, \qquad &&G^-_{+\frac{1}{2}} = \mathcal S^{2+}\;, \qquad &&\bar G^+_{+\frac{1}{2}}=\widetilde{\mathcal S}_{1}^{\dot +}\;, \qquad &&\bar G^-_{+\frac{1}{2}}=\mathcal S^{1-}\;.
\end{aligned}
\end{equation}
The identification of the bosonic generators $L_0,L_{\pm 1}, J_0$ and their barred versions follow immediately by matching the anticommutation relations among the supercharges. In particular, one finds for the two-dimensional vector and axial R-symmetries
\begin{equation}\label{RandAsymmetry}
R_V = J_0 + \bar J_0 = r\;,\qquad R_A = J_0-\bar J_0 = -4R - 2\mathcal M^\bot\;.
\end{equation}
It is important to remark that the supercharges $\qq_1$ and $\qq_2$ of section \ref{section:review-chiral-algebra} (see in particular footnote \ref{Qforcohomology}) are contained in the embedded two-dimensional algebra. This implies that an $\mathcal N=(2,2)$ preserving surface defect, transverse to the chiral algebra plane, participates in the cohomological construction of section \ref{section:review-chiral-algebra}, as claimed there already. Similarly, after performing a Weyl transformation, it means that our localization computation can be enriched by inserting an $\mathcal N=(2,2)$ superconformal surface defect along $S^2_{\theta=0}$. (See below for more details.) Also note that $R+\mathcal M^\bot$ is central to the embedding.

Finally, by analyzing the supersymmetry variations, it is easy to verify that the four-dimensional hypermultiplet scalars $q_{1A}$, for $A=1,2$, are the bottom components of two-dimensional \textit{twisted} chiral multiplets, and thus the hypermultiplet decomposes into a pair of twisted chiral multiplets (and their complex conjugate twisted anti-chiral multiplets).\footnote{Recall that the bottom component of the two-dimensional $\mathcal N=(2,2)$ supersymmetric multiplets in the left column of the following table are annihilated by the (Poincar\'e) supercharges in the right column
\renewcommand{\arraystretch}{1.3}
\begin{center}
\begin{tabular}{r|l}
multiplet & supercharges annihilating\\
& bottom component\\
\hline
chiral & $G^-_{-\frac{1}{2}},\, \bar G^-_{-\frac{1}{2}}$ \\
anti-chiral & $G^+_{-\frac{1}{2}},\, \bar G^+_{-\frac{1}{2}}$ \\
twisted chiral & $G^-_{-\frac{1}{2}},\, \bar G^+_{-\frac{1}{2}}$ \\
twisted anti-chiral & $G^+_{-\frac{1}{2}},\, \bar G^-_{-\frac{1}{2}}$ \\
\end{tabular}
\end{center}
\renewcommand{\arraystretch}{1.0}
Here we have omitted the less standard semi-chiral multiplets.}\textsuperscript{,}\footnote{The four-dimensional $\mathcal N=2$ vector multiplet decomposes in a two-dimensional $\mathcal N=(2,2)$ twisted vector multiplet and a twisted chiral multiplet.}

\subsection{4d/2d coupled system on \texorpdfstring{$S^4$}{S4}}
Returning to our setup on the four-sphere, we are interested in describing the defect as a 4d/2d coupled system. As explained above, the two-dimensional theory resides on $S^2_{\theta=0}$. It is useful to recall that a massive two-dimensional $\mathcal N=(2,2)$ supersymmetric theory can be placed on the two-sphere while preserving the symmetry algebra $\mathfrak{su}(2|1)$. Its $\mathfrak{su}(2)\oplus \mathfrak{u}(1)$ bosonic subalgebra consists of the spatial isometries of the two-sphere $\mathfrak{su}(2)\simeq\mathfrak{so}(3)$ and a $\mathfrak{u}(1)$ subalgebra of the vector and axial R-symmetry. The background preserving $\mathfrak{u}(1)_V$ is called the A-background, while the one preserving $\mathfrak{u}(1)_A$ is called the B-background \cite{Doroud:2012xw,Doroud:2013pka}. Momentarily, we will show that the localizing supercharge $\mathcal Q$ defined in \eqref{subsection:choiceofKS} resides in $\mathfrak{su}(2|1)_A$. One preliminary indication that this is the case is the fact that its square \eqref{Qsquared} equals a sum of a rotation of $S^2_{\theta=0}$ and, as can be read off from \eqref{RandAsymmetry}, the vector R-symmetry. Consequentially, our localization computation can proceed in the absence of a conformal UV Lagrangian description of the defect theory.

To verify the above statement, we show that the reduction of the four-dimensional Killing spinors $\xi_{I\alpha}, \tilde \xi_I^{\dot\alpha}$ of \eqref{Killing-spinor-solutions1}-\eqref{Killing-spinor-solutions2} to $S^2_{\theta=0}$ carry the vector and axial R-symmetry charges and satisfy the two-dimensional Killing spinor equation relevant for the $\mathfrak{su}(2|1)_A$ background. As explained in appendix \ref{subsection:reducespinorsontothetazero}, the reduction of arbitrary four-dimensional two-component chiral and anti-chiral spinors $\psi_\alpha$, $\tilde \psi^{\dot\alpha}$ to $S^2_{\theta = 0}$ is given by\footnote{Recall that the vielbein on $S^2_{\theta = 0}$ are induced from those on the four-sphere as in \eqref{vielbeinsS20} and that our gamma-matrices are chosen as in footnote \ref{footnote2dgamma}.} 
\begin{equation}
\psi \to \psi ^{{\text{2d}}} = \frac{1}{{\sqrt 2 }}\left( {\begin{array}{*{20}{c}}
    {{\psi _1} + {\psi _2}} \\ 
    {{\psi _1} - {\psi _2}} 
  \end{array}} \right),\qquad \tilde \psi \to  {\tilde \psi ^{{\text{2d}}}} = \frac{1}{{\sqrt 2 }}\left( {\begin{array}{*{20}{c}}
    {{{\tilde \psi }^1} - {{\tilde \psi }^2}} \\ 
    {{{\tilde \psi }^1} + {{\tilde \psi }^2}} 
  \end{array}} \right)\;.
\end{equation}
Applied to the Killing spinors $\xi_{I\alpha}, \tilde \xi_I^{\dot\alpha}$, we find (two-dimensional) (anti-)chiral spinors. We define $\epsilon, \tilde \epsilon$ by
\begin{equation}
  \xi _{I = 1} \to 2 {P_ - }\epsilon ,\qquad {\xi _{I = 2}} \to  - 2 {P_ + }\epsilon ,\qquad {\tilde \xi _{I = 1}} \to 2 {P_ + }\tilde \epsilon ,\qquad {\tilde \xi _{I = 2}} \to 2{P_ - }\tilde \epsilon\;.
\end{equation}
Here $P_{\pm}=\frac{1}{2}(1+\gamma_3)$ is the (two-dimensional) projection operator onto chiral/anti-chiral spinors. These definitions ensure that $\epsilon_{\pm}\equiv P_{\pm}\epsilon$ both carry one unit of $U(1)_V$ charge and have opposite $U(1)_A$ charge, while $\tilde \epsilon_{\pm}$ both have $U(1)_V$ charge minus one and once again opposite $U(1)_A$ charge. Moreover, the two-dimensional spinors can be shown to satisfy the equations
\begin{equation}
  \nabla^\text{2d} _\mu \epsilon = \frac{1}{{2r}}{\gamma _\mu }{\gamma _3}\epsilon ,\qquad {\nabla^\text{2d} _\mu }\tilde \epsilon  =  - \frac{1}{{2r}}{\gamma _\mu }{\gamma _3}\tilde \epsilon \;.
  \label{Killing-spinors-equation-2d}
\end{equation}
The $U(1)_V$ and $U(1)_A$ charge assignments of the two-dimensional Killing spinors and the equations they satisfy together prove that they indeed define a supercharge in $\mathfrak{su}(2|1)_A$. Note that the Killing spinors $\epsilon$ and $\tilde \epsilon$ precisely describe the supercharge used to localize two-dimensional $\mathcal N=(2,2)$ theories (in the A-background) on the two-sphere\cite{Benini:2012ui,Doroud:2012xw,Gomis:2012wy,Doroud:2013pka,Benini:2015isa}.

Before starting to analyze the 4d/2d coupled system, we provide the detailed decomposition of the four-dimensional hypermultiplet in a pair of two-dimensional twisted chiral multiplets $\Theta_A = (\phi_{\Theta_A},\tilde\phi_{\Theta_A}, \eta_{\Theta_A},\tilde \eta_{\Theta_A}, G_{\Theta_A}, \tilde G_{\Theta_A})$:
\begin{align}
&\begin{aligned}
&\phi_{\Theta_A}  \equiv {q_{1A}}\;,\qquad &&\tilde \phi_{\Theta_A}  \equiv  - {q_{2A}}\;,\\
&\eta_{\Theta_A}  \equiv 2\tilde \psi _A^{{\text{2d}}}\;, \qquad &&\tilde \eta_{\Theta_A}  \equiv 2{\gamma _3}\psi _A^{{\text{2d}}}\;,
\end{aligned}\nn\\
& G_{\Theta_A} \equiv  - ({D_2} - i{D_3})\tilde \phi_{\Theta_A}  + i\cot \frac{\rho }{2}{F_{1A}} \nn\\
& \tilde G_{\Theta_A} \equiv  - ({D_2} + i{D_3})\phi_{\Theta_A}  - i\tan \frac{\rho }{2}{F_{2A}} \;.\label{HMastCM}
\end{align}
One can verify that the four-dimensional supersymmetry transformations of these fields, when restricted to $S^2_{\theta=0}$, indeed precisely reproduce those of twisted chiral multiplets of Weyl weight $\Delta_{\Theta_A} = 1$ placed on a two-sphere. The latter are, for example, given in formula 4.3 of \cite{Gomis:2012wy}.\footnote{Note that the notations $\eta$ and $\tilde \eta$ are slightly misleading since the fields with and without tilde do not separate in multiplets. Rather, $(\phi, \eta_-, \tilde \eta_+, G)$ form a twisted chiral multiplet, while $(\tilde \phi, \tilde \eta_-, \eta_+, \tilde G)$ form a twisted anti-chiral multiplet.} We emphasize that $D_2$ and $D_3$ (are proportional to) derivatives along the $\chi$ and $\theta$ directions, which are orthogonal to the two-sphere $S^2_{\theta = 0}$.

In the absence of four-dimensional gauge fields, the coupling between the four-dimensional and two-dimensional degrees of freedom takes place via superpotentials. More precisely, since the free hypermultiplet decomposes in a pair of twisted chiral multiplets, such coupling is implemented by a twisted superpotential. Let the native two-dimensional degrees of freedom be described by some theory $\mathcal T_{\text{2d}}$ involving vector multiplets, twisted vector multiplets, chiral multiplets and twisted chiral multiplets.\footnote{We do not consider the less standard semi-chiral multiplets.} Let $\Lambda = (\phi, \tilde \phi, \eta, \tilde \eta, G, \tilde G)_\Lambda$ collectively denote the twisted chiral multiplets and $\Sigma = ({\sigma _2} + i{\sigma _1},{\sigma _2} - i{\sigma _1},\lambda,\tilde \lambda,D - \frac{\sigma_2}{r} + iF_{12},D - \frac{\sigma_2}{r} - iF_{12})$ the twisted chiral field strength multiplet. The most general twisted superpotential describing the coupling between $\mathcal T_{\text{2d}}$ and the four-dimensional free hypermultiplet, as well as any additional native two-dimensional twisted superpotential couplings, is then described by a holomorphic function $\widetilde W(\phi_\Sigma, \phi_\Lambda, \phi_\Theta)$. It enters in the Lagrangian as \cite{Gomis:2012wy}
\begin{equation}\label{twistedsuperpotentialcoupling}
  \mathcal{L}_\text{tc} = \mathcal{L}_\text{tc}^{\text{2d}} + \mathcal{L}_\text{tc}^{\text{4d/2d}}  = \Bigg[ {\frac{i}{r}\widetilde W(\phi ) - i\frac{{\partial \widetilde W}}{{\partial {\phi_j}}}\Big( {{G_j} + \frac{{{\Delta_j}}}{r}{\phi_j}} \Big) - \frac{{{\partial ^2}\widetilde W(\phi )}}{{\partial {\phi_j}\partial {\phi _k}}}({{\tilde \eta }_k}{P_ - }\eta_j )} \Bigg] + \text{c.c}\;.
\end{equation}
Here $\mathcal{L}_\text{tc}^{\text{2d}}$ denotes the Lagrangian describing the twisted superpotential couplings among native two-dimensional degrees of freedom only, while $\mathcal{L}_\text{tc}^{\text{4d/2d}}$ are all other terms in $\mathcal{L}_\text{tc}$, \ie{}, the terms describing the coupling between the four-dimensional and two-dimensional degrees of freedom.\footnote{Note that if one were to write the four-dimensional free hypermultiplet action in terms of its reduced twisted chiral multiplets on the two-sphere, additional twisted superpotential terms would occur.} Furthermore, the indices $j,k,\ldots$ run over $\Sigma, \Lambda, \Theta$ (and their subindices) and are summed over in the standard way, $\Delta$ denotes the Weyl weight of the respective multiplet and the complex conjugation simply replaces the function $\widetilde W(\phi_\Sigma, \phi_\Lambda, \phi_{\Theta_A})$ by $\widetilde W(\tilde \phi_\Sigma, \tilde \phi_\Lambda, - \epsilon^{AB} \tilde \phi_{\Theta_B})$.

The total action of the 4d/2d coupled system of interest in this paper is thus
\begin{equation}
S_{\text{total}} = S_{\text{HM}}^{S^4} + S_{\mathcal T_{\text{2d}}}^{S^2_{\theta=0}} + \int_{S^2_{\theta=0}} d^2x \sqrt{g_{S^2}}\ \mathcal L_{\text{tc}}^{\text{4d/2d}}\;.
\end{equation}
For localization purposes, it is important to note that the kinetic terms for native two-dimensional vector multiplets, twisted vector multiplets, chiral multiplets and twisted chiral multiplets are all $\mathcal Q$-exact, and so are any superpotential couplings. The Lagrangian $\mathcal{L}_\text{tc}$ of \eqref{twistedsuperpotentialcoupling} describing the twisted superpotential couplings, however, is $\mathcal Q$-closed but not exact.

\subsection{Localization of 4d/2d coupled system on \texorpdfstring{$S^4$}{S4}}
To perform a localization computation of the 4d/2d coupled system, one starts by localizing the theory $\mathcal T_{\text{2d}}$ describing the dynamics of the native two-dimensional degrees of freedom. Since, as remarked in the above subsection, the supercharge with respect to which the localization computation is to be performed is the standard choice in $\mathfrak{su}(2|1)_A$, the results of \cite{Benini:2012ui,Doroud:2012xw,Gomis:2012wy,Doroud:2013pka} are directly applicable.\footnote{The localization of non-abelian twisted vector multiplets has not yet appeared in the literature.} We refer the readers to these papers for all details on the computation. 

It is relevant to stress that the classical action evaluated on the BPS configurations receives only nontrivial contributions from the twisted superpotential terms, as all other actions are $\mathcal Q$-exact. In the computation at hand, one finds in particular a non-zero contribution from $\int_{S^2_{\theta=0}} d^2x \sqrt{g_{S^2}}\ \mathcal L_{\text{tc}}^{\text{4d/2d}}$. To obtain a detailed understanding of the effect of the 4d/2d twisted superpotential coupling, we summarize the BPS configurations of multiplets that enter $\widetilde W$:
\begin{align}
&\text{twisted chiral:}\quad &&\phi_\Lambda = \text{const}\;, \qquad &&\tilde \phi_\Lambda = \text{const}\;, \qquad &&{G_\Lambda } + \frac{{{\Delta _\Lambda }}}{r}{\phi _\Lambda } = {{\tilde G}_\Lambda } + \frac{{{\Delta _\Lambda }}}{r}{{\tilde \phi }_\Lambda } = 0\;,\nn\\
&\text{vector:}\quad &&{\sigma _1} = - \frac{B}{{2r}}\;,\qquad &&{\sigma _2} = \text{const}\;,\qquad &&F_{12} = \frac{B}{2r^2}\;, \qquad D= 0 \;,\label{2d-BPS-configs}
\end{align}
where $B$ is a GNO quantized (constant) magnetic flux through the two-sphere. We thus find for the evaluation of $\mathcal L_{\text{tc}}$ on the two-dimensional localization locus
\begin{equation}
{\mathcal{L}_{{\text{tc}}}} = \Bigg[ \frac{i}{r}\widetilde W(\phi ) - i\frac{{\partial \widetilde W}}{{\partial {\phi _\Theta }}}({G_\Theta } + \frac{1}{r}{\phi _\Theta }) - \frac{{{\partial ^2}\widetilde W(\phi )}}{{\partial {\phi _\Theta }\partial {\phi _\Theta }}}({\tilde \eta _\Theta }{P_ - }{\eta _\Theta }) \Bigg] + \text{c.c}\;.\label{tcLafter2dloc}
\end{equation}
Here $\widetilde W(\phi)$ is still a holomorphic function of the scalar bottom components of all twisted chiral multiplets $\Sigma, \Lambda, \Theta$, but with $\phi_\Lambda$ and $\phi_\Sigma$ set to constants as in \eqref{2d-BPS-configs}. Note also that we used that the Weyl weights $\Delta_{\Theta_A}$ of the twisted chiral multiplets $\Theta_A$ obtained from the reduction of the free hypermultiplet equal one.

As a next step, one performs the localization of the four-dimensional free hypermultiplet. The computation is identical to the one in section \ref{section:freeHM}, except for that we additionally need to evaluate the twisted superpotential 4d/2d coupling. To do so, we first remark that all fermions in the hypermutliplet are set to zero, and then observe that the BPS equations \eqref{BPS-equations} imply that at the locus $\theta=0$ the combinations \eqref{HMastCM} satisfy
\begin{equation}
\text{at }\theta=0:\quad {G_{{\Theta _A}}}\, \sin \rho = + \frac{1}{r} {\partial_\rho }((1 + \cos \rho )\phi_{\Theta_A} )\;, \quad {{\tilde G}_{\Theta_A}}\, \sin \rho  = - \frac{1}{r}{\partial_\rho }((1 - \cos \rho ){{\tilde \phi }_{{\Theta _A}}} \;.
\end{equation}
It is then straightforward to verify that the Lagrangian term \eqref{tcLafter2dloc} multiplied with the integration measure simplifies to a total $\rho$-derivative:
\begin{align}
  \sqrt {{g_{S_{\theta  = 0}^2}}} \left[ {\frac{i}{r}\widetilde W(\phi ) - i\frac{{\partial \widetilde W}}{{\partial {\phi _\Theta }}}({G_\Theta } + \frac{1}{r}{\phi _\Theta })} \right] = & - \frac{i}{r}{\partial _\rho }\big((1 + \cos \rho )\widetilde W(\phi )\big) \nn\\
  \sqrt {{g_{S_{\theta  = 0}^2}}} \left[ {\frac{i}{r}\widetilde W(\tilde \phi ) - i\frac{{\partial \widetilde W}}{{\partial {{\tilde \phi }_\Theta }}}({{\tilde G}_\Theta } + \frac{1}{r}{{\tilde \phi }_\Theta })} \right] = & + \frac{i}{r}{\partial _\rho }\big((1 - \cos \rho )\widetilde W(\tilde \phi )\big)\nn\;.
\end{align}
Note that here we used that the profiles of the native two-dimensional fields entering in the twisted superpotential is constant. Recalling that on the localization locus both two-dimensional and four-dimensional fields are independent of $\varphi$, it is trivial to perform the integral over $S^2_{\theta=0}$. We find
\begin{equation}
{S_{{\text{tc}}}} = \int_0^{2\pi } {d\varphi \int_0^\pi  {d\rho \sqrt {{g_{S_{\theta  = 0}^2}}} {\mathcal{L}_{{\text{tc}}}}} }  =   4\pi ir \Big[\widetilde W({\phi _\Theta }({\text{NP}}),\phi_\Lambda,\phi_\Sigma) + \widetilde W({{\tilde \phi }_\Theta }({\text{SP}}),\tilde\phi_\Lambda,\tilde\phi_\Sigma) \Big]\;,
\end{equation}
where NP denotes the north pole $\rho=0$ and SP the south pole $\rho=\pi$. As before $\phi_\Lambda$ and $\phi_\Sigma$ are set to constants as in \eqref{2d-BPS-configs}. It is also useful to recall that
\begin{equation}
\phi _{\Theta _A}(\text{NP}) = Q^\text{(N)}_A(\text{NP}),\qquad \tilde \phi _{\Theta _A}(\text{SP}) = - i Q^\text{(S)}_A(\text{SP})\;.
\end{equation}

In summary, the partition function of our 4d/2d coupled system is given by
\begin{equation}
Z_{\text{4d/2d}} = \SumInt [d\phi_{\text{2d}}]\ Z_{\text{1-loop}}(\phi_{\text{2d}}) \int [DQ D\widetilde Q]\ e^{-S^\text{2d}_{{\text{HM}}}[Q,\widetilde Q]} \ e^{- 4\pi ir \big[\widetilde W({\phi _\Theta }({\text{NP}}),\phi_\Lambda,\phi_\Sigma) + \widetilde W({{\tilde \phi }_\Theta }({\text{SP}}),\tilde\phi_\Lambda,\tilde\phi_\Sigma) \big]}\;.
\end{equation}
Here $\phi_{\text{2d}}$ collectively denotes the localization locus of the two-dimensional theory $\mathcal T_{\text{2d}}$. The one-loop determinant of quadratic fluctuations of the two-dimensional theory is denoted $Z_{\text{1-loop}}(\phi_{\text{2d}})$. Once again, in the twisted superpotential, $\phi_\Lambda$ and $\phi_\Sigma$ are set to constants as in \eqref{2d-BPS-configs}.\footnote{Note that these constant profiles are integrated over by the integral $\SumInt [d\phi_{\text{2d}}]$.} Note that the localization computation can be extended to compute correlators also for the coupled system. As in subsection \ref{subsection:finalResultHMlocalization} the computation is compatible with the insertion of composites of $Q$ and $\widetilde Q$ on $S^2_{\theta=\frac{\pi}{2}}$. Moreover, we can also insert two-dimensional $\mathcal Q$-closed observables. In particular, polynomials in the scalar fields $\phi_\Lambda, \phi_{\Sigma}$ and $\tilde \phi_\Lambda,\tilde \phi_{\Sigma}$ of the twisted chiral multiplets $\Lambda,\Sigma$ can be inserted at the north pole and south pole respectively. It may be useful to remark that the twisted superpotential can be reinterpreted as the insertion of a particular pair of mixed 4d/2d observables $\widetilde V_{\text{NP/SP}}$ at the north pole and south pole:
\begin{equation}
\widetilde V_{\text{NP}}\equiv e^{- 4\pi ir \widetilde W(\phi _\Theta(\text{NP}),\phi_\Lambda,\phi_\Sigma) }\;, \qquad \widetilde V_{\text{SP}}\equiv e^{ - 4\pi ir  \widetilde W({\tilde \phi_\Theta }(\text{SP}),\tilde\phi_\Lambda,\tilde\phi_\Sigma)}\;.
\end{equation}
Also note that the additional insertions of two-dimensional observables at the poles of $S^2_{\theta=\frac{\pi}{2}}$ are a crucial ingredient to build the module of the chiral algebra, as explained at the end of section \ref{section:review-chiral-algebra}, and as we will explore in more detail in the next subsection.

\subsection{Computing structure constants of chiral algebra modules}
In section \ref{section:review-chiral-algebra}, we reviewed that the insertion of a surface defect -- orthogonal to the chiral algebra plane and preserving $\mathcal N=(2,2)$ supersymmetry on its worldvolume -- translates to considering a module $M$ of the chiral algebra. The localization results of the previous subsection allow us to analyze this module quantitatively. 

For the class of 4d/2d coupled systems under consideration, the space $M$ is easily identified. Its basis $e_i$ is given by all words built from the elementary letters $Q,\widetilde Q, \phi_\Lambda, \phi_\Sigma$ and $\wp$. The latter acts on $Q$ and $\widetilde Q$ only. Precisely these combinations can be inserted at the north pole in the localized path integral. Some care has to be taken when inserting composites of $Q,\widetilde Q$ at the north pole, as there already is an insertion of $\widetilde V_{\text{NP}}$ present there. We regularize the resulting divergence by subtracting off the contractions of the composite and $\widetilde V_{\text{NP}}$. Let us next define a dual space $M^\star$. Its basis $e^i$ comprises all words strung from the elementary letters $Q,\widetilde Q, \tilde\phi_\Lambda, \tilde\phi_\Sigma$ and $\wp$ inserted at the south pole. We can define a sesquilinear form
\begin{equation}
\langle \cdot , \cdot \rangle: M^\star\times M \rightarrow \mathbb C
\end{equation} 
defined by
\begin{equation}\label{ingr1}
\langle e^j , e_i \rangle = \frac{1}{Z_{\text{4d/2d}}} \SumInt [d\phi_{\text{2d}}]\ Z_{\text{1-loop}}(\phi_{\text{2d}}) \int [DQ D\widetilde Q]\ \left(e^j\, \widetilde V_{\text{SP}}\right) \left(e_i\, \widetilde V_{\text{NP}}\right) e^{-S^\text{2d}_{{\text{HM}}}[Q,\widetilde Q]} \;.
\end{equation}
and extended by sesquilinearity. Here the brackets around $e_i\, \widetilde V_{\text{NP}}$ denote the regularization mentioned above, and similarly for $e^j\, \widetilde V_{\text{SP}}$. In other words, no contractions should be considered within the bracketed combinations. It will be useful to introduce the notation $N_i^{\phantom{i}\, j} \equiv \langle e^j , e_i \rangle$. We will assume that $N$ is invertible.

The algebra acts on the module $M$ as explained around \eqref{algebraonmodule}. We would like to compute the structure constants defined in \eqref{defstructureconstants}. Using the path integral it is more convenient to consider the action of the Schur operator $a(\rho,\chi)$ as a field. The action of individual modes can afterwards be easily extracted. On the basis element $e_i$, we thus have
\begin{equation}
a(\rho,\chi)\cdot e_i = \sum_k \lambda_{ai}^{\phantom{ai}k}(\rho,\chi)\ e_k \;.
\end{equation}
Taking the pairing with $e^j$, one finds
\begin{equation}
\langle e^j , a(\rho,\chi)\cdot e_i \rangle =  \sum_k \lambda_{ai}^{\phantom{ai}k}(\rho,\chi)\ \langle e^j , e_k \rangle = \sum_k \lambda_{ai}^{\phantom{ai}k}(\rho,\chi)\ N_k^{\phantom{k}\, j}\;.
\end{equation}
Isolating the structure constants is then, in principle, straightforward:
\begin{equation}\label{isolatedlambda}
\lambda_{ai}^{\phantom{ai}k}(\rho,\chi) = \sum_j (N^{-1})_j^{\phantom{j}\, k}\langle e^j , a(\rho,\chi)\cdot e_i \rangle\;.
\end{equation}
Note that the right-hand side of this expression can be computed explicitly from the path integral using \eqref{ingr1} and 
\begin{equation}\label{ingr2}
\langle e^j , a(\rho,\chi)\cdot e_i \rangle = \frac{1}{Z_{\text{4d/2d}}} \SumInt [d\phi_{\text{2d}}]\ Z_{\text{1-loop}}(\phi_{\text{2d}}) \int [DQ D\widetilde Q]\ \left(e^j\, \widetilde V_{\text{SP}}\right) a(\rho,\chi) \left(e_i\, \widetilde V_{\text{NP}}\right) e^{-S^\text{2d}_{{\text{HM}}}[Q,\widetilde Q]}\;.
\end{equation}

The expression \eqref{isolatedlambda} provides a concrete tool to verify the dependence of the module on coupling constants. (See footnote \ref{footnoteparamdependence} for the definition of when a module is said to depend on a parameter.) Unfortunately, in practice, evaluating \eqref{isolatedlambda} is not a computation that can be finished in finite time. It seems quite unlikely though that for any choice of two-dimensional degrees of freedom $\mathcal T_{\text{2d}}$ and any choice of twisted superpotential coupling the native two-dimensional degrees of freedom to the four-dimensional ones, the parameter dependence would drop out. In other words, generically, one would expect the module to depend on coupling constants. It would be very interesting to find a way to make this statement concrete, either in an example, or abstractly.


\section{Discussion and future directions}
In this paper we have used supersymmetric localization techniques to show that the chiral algebra associated with four-dimensional $\mathcal N=2$ superconformal quantum field theories is accessible directly in the path integral. We did so for the specific case of the free hypermultiplet. We extended the localization computation to include surface defects and showed that this point of view provides a computational handle on properties of the resulting chiral algebra modules not easily accessible by other methods. Many future directions present themselves. We list a few of the most promising ones:
\begin{itemize}
\item In this paper we have focused on the four-dimensional free hypermultiplet. Extending the computation to also include vector multiplets is an obvious next step. It is relatively straightforward to convince oneself that the vector multiplet BPS locus is not purely bosonic. In other words, the deformation action admits fermionic zero-modes around the bosonic BPS configurations. Taking these fermionic zero-modes into account properly presumably works along the lines of \cite{Benini:2013nda,Benini:2013xpa,Benini:2015noa,Closset:2015rna}. The classical action for the vector multiplet, when evaluated on the solutions of the (complex) BPS equations, can be shown to reduce to an action on the two-sphere $S^2_{\theta=\frac{\pi}{2}}$, as can the gauged hypermultiplet action. We present these computations in appendix \ref{app:evaluateCA}. Finally, one should evaluate the one-loop determinants. 

Note that in the absence of any operator insertions, one is simply computing the four-sphere partition function and should thus reproduce the results of \cite{Pestun:2007rz}. It would be particularly fascinating to see how the instanton contributions are captured in this alternative localization computation. An intriguing, and possibly relevant, observation made in \cite{Nekrasov:2003rj} is that the dual instanton partition function has a description in terms of correlators constructed from chiral fermions.

\item The structure of a chiral algebra is naturally quite rigid. Nevertheless, one should study if and to what extent the computation presented in this paper can be deformed, either geometrically, by deforming the four-sphere, or field theoretically, by turning on BPS configurations for various background fields. Note that deformations of the latter kind were observed to be feasible in a similar computation on the three-sphere \cite{Dedushenko:2016jxl}. In appendix \ref{appendix_oneloop} we consider the effect on the partition function of an example of such a deformation. Moreover, if the localization computation goes through on the squashed four-sphere, it would be extremely interesting to consider the non-canonical deformation action used in the Higgs branch localization computations of \cite{Chen:2015fta,Pan:2015hza}.
\item In this paper we have considered the insertion of defects orthogonal to the chiral algebra sphere. To be compatible with the localizing supercharge, these defects had to preserve two-dimensional $\mathcal N=(2,2)$ supersymmetry. It is easy to verify that the localizing supercharge is also compatible with the insertion of defects on the chiral algebra sphere if they preserve two-dimensional $\mathcal N=(0,4)$ supersymmetry. The insertion of such defects is expected to modify the structure of the resulting two-dimensional theory nontrivially. In appendix \ref{subsec:singular-config}, we describe a simple singular profile for a (background) vector multiplet representing such a defect. One could also consider exploring systems of intersecting surface defects, as initiated in \cite{Gomis:2016ljm}. (See also \cite{Pan:2016fbl}.)
\item The four-sphere partition function of four-dimensional $\mathcal N=2$ theories of class $\mathcal S$ features prominently in the AGT correspondence \cite{Alday:2009aq}. It seems natural to expect that the new representation of the four-sphere partition function obtained in this paper, and in particular its extension to include vector multiplets mentioned above, will provide new perspectives on this correspondence and on its relationship to the chiral algebras associated to theories of class $\mathcal S$ \cite{Beem:2014rza,Lemos:2014lua}.
\end{itemize}
We hope to report on progress on some of these problems in the future.

\section*{Acknowledgments}
The authors are grateful to Chris Beem, Guido Festuccia, Jaume Gomis, Pietro Longhi, Leonardo Rastelli, Balt van Rees, Martin Rocek, Yifan Wang and Maxim Zabzine for helpful conversations and useful suggestions. Y.P. is supported in part by Vetenskapsr{\aa}det under grant {\#}2014- 5517, by the STINT grant and by the grant ``Geometry and Physics'' from the Knut and Alice Wallenberg foundation. The work of W.P. is supported in part by the DOE grant DOE-SC0010008.

\appendix

\section{Reduction of spinors}
In section \ref{section:freeHM} we showed that the theory of the free hypermultiplet reduces to a quantum field theory defined on the submanifold $S^2_{\theta=\frac{\pi}{2}}$, while in section \ref{section:defects} we studied the coupling of the four-dimensional theory to native degrees of freedom on the submanifold $S^2_{\theta=0}$. In this appendix we elaborate on an important technical step in these computations, namely the reduction of four-dimensional spinors to these two-spheres.

In general, the first step to reduce spinors onto a two-dimensional submanifold $\Sigma \subset S^4$ is to find a suitable similarity transformation $U_\Sigma$ that casts the four-dimensional $\Gamma$-matrices in a useful Kronecker product form. Let $E^{A = 1,2,3,4}$ denote the vielbein on the four-sphere $S^4$. Without loss of generality, we assume that the restriction to the submanifold $\Sigma$ is such that $E^\alpha|_\Sigma = e^1$, $E^\beta|_\Sigma = e^2$, for some $\alpha,\beta$, where $e^{a = 1,2}$ are vielbein on $\Sigma$. We then look for a similarity transformation $U_\Sigma$ such that
\begin{equation}
\Gamma^\text{new}_\alpha = {U_\Sigma}{\Gamma _\alpha}U_\Sigma^{ - 1} = {\gamma _1} \otimes (\ldots),\qquad \Gamma^\text{new}_\beta = {U_\Sigma}{\Gamma _\beta}U_\Sigma^{ - 1} = {\gamma _2} \otimes (\ldots)\;, \qquad \ldots\;.
\end{equation}
Here the $(\ldots)$ are two-dimensional matrices. Note that the choice of $U_\Sigma$ is not unique: the freedom can be exploited to obtain maximal simplicity. Let us then consider any four-component spinor $\Psi$ on $S^4$. When acted on by the transformation $U_\Sigma$, it can be written in a tensor product form $U_\Sigma\Psi = \psi^\text{2d} \otimes (\ldots) + \tilde \psi^\text{2d} \otimes (\ldots)$, where $(\ldots)$ are simple, constant two-component columns. The action of four-dimensional gamma-matrices on the spinor $\Psi$ can be straightforwardly reduced to an action on the spinors $\psi^\text{2d},\tilde\psi^\text{2d}$. Let us apply this logic to the two cases of interest.

\subsection{Reduction onto \texorpdfstring{$\Sigma = S^2_{\theta = 0}$}{S2}}\label{subsection:reducespinorsontothetazero}

Let us first consider the reduction onto $S^2_{\theta = 0}$. The four-dimensional vielbein reduce as $E^4 \to e^1$, ${E^1} \to e^2$, where ${e^1} = rd\rho $, $e^2 = r \sin\rho d\varphi$ are the vielbein on $S^2_{\theta = 0}$. We choose the transformation matrix $U_{\theta = 0}$ to be
\begin{equation}
  {U_{\theta  = 0}} = \frac{1}{{\sqrt 2 }}\left( {\begin{array}{*{20}{c}}
    1&1&0&0 \\ 
    0&0&1&{ - 1} \\ 
    1&{ - 1}&0&0 \\ 
    0&0&1&1 
  \end{array}} \right)\;,\qquad
  \begin{array}{*{20}{l}}
    {\Gamma _{1}^{{\text{new}}} = {\gamma _2} \otimes {\gamma _1}} \\ 
    {\Gamma _{4}^{{\text{new}}} = {\gamma _1} \otimes {\gamma _1}} \\ 
    {\Gamma _{2}^{{\text{new}}} = {\gamma _3} \otimes {\gamma _1}} \\ 
    {\Gamma _{3}^{{\text{new}}} = {\unit_{2 \times 2}} \otimes {\gamma _2}} 
  \end{array}
\end{equation}
In this new representation, a four-component spinor $\Psi = (\psi_\alpha, \tilde \psi^{\dot \alpha})$ on $S^4$ is rotated into
\begin{equation}
  \Psi^\text{new} = U_{\theta  = 0}\Psi  = \frac{1}{{\sqrt 2 }}\left( {\begin{array}{*{20}{c}}
    {{\psi _1} + {\psi _2}} \\ 
    {{{\tilde \psi }^1} - {{\tilde \psi }^2}} \\ 
    {{\psi _1} - {\psi _2}} \\ 
    {{{\tilde \psi }^1} + {{\tilde \psi }^2}} 
  \end{array}} \right) = {\psi ^{{\text{2d}}}} \otimes \left( {\begin{array}{*{20}{c}}
    1 \\ 
    0 
  \end{array}} \right) + {{\tilde \psi }^{{\text{2d}}}} \otimes \left( {\begin{array}{*{20}{c}}
    0 \\ 
    1 
  \end{array}} \right)\;,
\end{equation}
where ${\psi ^{{\text{2d}}}} \equiv \frac{1}{\sqrt 2}{\left( {{\psi _1} + {\psi _2},{\psi _1} - {\psi _2}} \right)^T}$, ${{\tilde \psi }^{{\text{2d}}}} \equiv \frac{1}{\sqrt 2}{( {{{\tilde \psi }^1} - {{\tilde \psi }^2},{{\tilde \psi }^1} + {{\tilde \psi }^2}} )^T}$.

The induced action of the $\sigma$- and $\tilde \sigma$-matrices (or equivalently the four-dimensional $\Gamma$-matrices) on the two-dimensional spinors $\psi^\text{2d}$ and $\tilde \psi^\text{2d}$ is simply
\begin{align}
  & {\sigma _1}\tilde \psi  \to \gamma _2{\tilde \psi }^\text{2d},\quad {\sigma _2}\tilde \psi  \to \gamma _3\tilde \psi ^\text{2d},\quad {\sigma _3}\tilde \psi  \to - i \tilde \psi ^\text{2d},\quad {\sigma _4}\tilde \psi  \to \gamma _1\tilde \psi ^\text{2d}\\
  & {{\tilde \sigma }_1}\psi  \to \gamma _2\psi ^\text{2d},\quad {{\tilde \sigma }_2}\psi  \to \gamma _3\psi ^\text{2d},\quad {{\tilde \sigma }_3}\psi  \to + i \psi ^\text{2d},\quad {{\tilde \sigma }_4}\psi  \to \gamma _1\psi ^\text{2d} \ .
\end{align}

The two-dimensional inner product of spinors is defined to be $(\psi^\text{2d}\chi^\text{2d}) \equiv (\psi^\text{2d})^T (i \tau_2) \chi$. It is related to the four-dimensional inner product between two-component undotted spinors $\psi$ and $\chi$ and between dotted spinors $\tilde \psi$ and $\tilde \chi$ by $(\psi \chi ) = ({\psi ^{{\text{2d}}}}{\chi ^{{\text{2d}}}})$, $(\tilde \psi \tilde \chi ) = ({{\tilde \psi }^{{\text{2d}}}}{{\tilde \chi }^{{\text{2d}}}})$.

\subsection{Reduction onto \texorpdfstring{$\Sigma = S^2_{\theta = \pi/2}$}{S2}}\label{subsection:reducespinorsontothetapiotwo}
Next, we consider the reduction onto $S^2_{\theta=\frac{\pi}{2}}$. The four-dimensional vielbein reduce as $E^4 \to {e^1}$, ${E^2} \to {e^2}$, where ${e^1} = rd\rho $, $e^2 = r \sin\rho d\chi$ are the vielbein on $S^2_{\theta=\frac{\pi}{2}}$ . We choose $U_{\theta = \frac{\pi}{2}}$ to be
\begin{equation}
  {U_{\theta  = \frac{\pi}{2}}} = \frac{1}{\sqrt 2}\left( {\begin{array}{*{20}{c}}
    1&{ - i}&0&0 \\ 
    0&0&{ - i}&1 \\ 
    { - i}&1&0&0 \\ 
    0&0&1&{ - i} 
  \end{array}} \right),\qquad \begin{array}{*{20}{l}}
    \Gamma _4^{{\text{new}}} = {\gamma _1} \otimes {\gamma _1} \\ 
    \Gamma _2^{{\text{new}}} = {\gamma _2} \otimes {\gamma _1} \\ 
    \Gamma _3^{{\text{new}}} = {\gamma _3} \otimes {\gamma _1} \\ 
    \Gamma _1^{{\text{new}}} = {\unit_{2 \times 2}} \otimes {\gamma _2} 
  \end{array}\;.
\end{equation}
In this new representation
\begin{equation}
  \Psi^\text{new} = U_{\theta  = \frac{\pi}{2}}\Psi  = \frac{1}{{\sqrt 2 }}\left( {\begin{array}{*{20}{c}}
    {{\psi _1} - i{\psi _2}} \\ 
    { - i{{\tilde \psi }^1} + {{\tilde \psi }^2}} \\ 
    { - i{\psi _1} + {\psi _2}} \\ 
    {{{\tilde \psi }^1} - i{{\tilde \psi }^2}} 
  \end{array}} \right) = {\psi ^{{\text{2d}}}} \otimes \left( {\begin{array}{*{20}{c}}
    1 \\ 
    0 
  \end{array}} \right) + {{\tilde \psi }^{{\text{2d}}}} \otimes \left( {\begin{array}{*{20}{c}}
    0 \\ 
    1 
  \end{array}} \right)\;,
\end{equation}
where ${\psi ^{{\text{2d}}}} = \frac{1}{\sqrt 2}{({\psi _1} - i{\psi _2}, - i{\psi _1} + {\psi _2})^T}$, ${{\tilde \psi }^{{\text{2d}}}} = \frac{1}{\sqrt 2}{( - i{{\tilde \psi }^1} + {{\tilde \psi }^2},{{\tilde \psi }^1} - i{{\tilde \psi }^2})^T}$. The action of the $\sigma$- and $\tilde \sigma$-matrices reduce as
\begin{align}
  & {\sigma _1}\tilde \psi  \to  - i{{\tilde \psi }^{{\text{2d}}}},\quad {\sigma _2}\tilde \psi  \to {\gamma _2}{{\tilde \psi }^{{\text{2d}}}},\quad {\sigma _3}\tilde \psi  \to {\gamma _3}{{\tilde \psi }^{{\text{2d}}}},\quad {\sigma _4}\tilde \psi  \to {\gamma _1}{{\tilde \psi }^{{\text{2d}}}}\\
  & {{\tilde \sigma }_1}\psi  \to  + i{\psi ^{{\text{2d}}}},\quad {{\tilde \sigma }_2}\psi  \to {\gamma _2}{\psi ^{{\text{2d}}}},\quad {{\tilde \sigma }_3}\psi  \to {\gamma _3}{\psi ^{{\text{2d}}}},\quad {{\tilde \sigma }_4}\psi  \to {\gamma _1}{\psi ^{{\text{2d}}}}\;.
\end{align}

In section \ref{section:review-chiral-algebra}, we encountered the spinorial combinations $\mathfrak{q}_A \equiv q{^I}_A \xi_I$, $\tilde{\mathfrak{q}}_A \equiv  q{^I}_A \tilde\xi_I$. Using the explicit expression for the Killing spinors \eqref{Killing-spinor-solutions1}-\eqref{Killing-spinor-solutions2}, the corresponding four-component spinor $(\mathfrak{q}_{A\alpha}, \tilde{\mathfrak{q}}_A^{\dot \alpha})$ reduces to
\begin{equation}
  (\mathfrak{q}_{A\alpha}, \tilde{\mathfrak{q}}^{\dot\alpha}_A) \to (i - 1)\; \mathfrak{q}_A^{{\text{2d}}} \otimes \left( {\begin{array}{*{20}{c}}
    1 \\ 
    0 
  \end{array}} \right) - (1 + i) \; \mathfrak{q}_A^{{\text{2d}}} \otimes \left( {\begin{array}{*{20}{c}}
    0 \\ 
    1 
  \end{array}} \right)\;,
  \label{reduction-of-qxi}
\end{equation}
where $\mathfrak{q}_A^{{\text{2d}}} \equiv {(0, e^{ - \frac{i\chi}{2}}\cos (\frac{\rho}{2}){q_{1A}} - i\lambda {e^{ + \frac{i\chi}{2} }}\sin (\frac{\rho}{2}){q_{2A}})^T}$ is an anti-chiral spinor on $S^2_{\theta = \frac{\pi}{2}}$.\footnote{This expression for $\mathfrak{q}^\text{2d}_A$ is valid in the patch $U_{\text{eq}}$ of the two-sphere, as that is the region of validity of the vielbein we used. See footnote \ref{footnotePatches} for more details on the patches and the frame rotations between them.} We note that, although we start with two pairs of spinors $\mathfrak{q}_A$, $\tilde {\mathfrak{q}}_A$ in four dimensions, only a single anti-chiral pair $\mathfrak{q}^\text{2d}_A$ appears in the reduction.

Again, the two-dimensional inner product of spinors is given by $(\psi^\text{2d}\chi^\text{2d}) \equiv (\psi^\text{2d})^T (i \tau_2) \chi$, and is related to the four-dimensional inner product between two-component undotted spinors $\psi$ and $\chi$ and between dotted spinors  $\tilde \psi$ $\tilde \chi$ as $(\psi \chi ) = - ({\psi ^{{\text{2d}}}}{\chi ^{{\text{2d}}}})$, $(\tilde \psi \tilde \chi ) = - ({{\tilde \psi }^{{\text{2d}}}}{{\tilde \chi }^{{\text{2d}}}})$.


\section{Vector multiplets and gauged hypermultiplets}\label{app:VM}
In the main text we performed the localization computation of free hypermultiplets in great detail. In this appendix we would like to collect some results on the localization of vector multiplets and gauged hypermultiplets. In appendix \ref{appendix_VMBPS} we analyze in some detail the BPS equations for the vector multiplet. Next, we evaluate the classical actions of the vector multiplet and gauged hypermultiplet on solutions of the complex BPS equations and show that they both reduce to an action on the two-sphere $S^2_{\theta=\frac{\pi}{2}}$. In appendix \ref{appendix_oneloop} we consider a particularly interesting solution to the vector multiplet BPS equations and compute the partition function of the resulting two-dimensional theories with this background turned on. Finally, in appendix \ref{subsec:singular-config} we analyze in some more detail a singular solution to the vector multiplet equations and show that it corresponds to an $\mathcal N=(0,4)$ preserving surface defect on $S^2_{\theta=\frac{\pi}{2}}$.

\subsection{BPS equations for the vector multiplet}\label{appendix_VMBPS}
We would like to study solutions to the vector multiplet BPS equations. The relevant equations were given in \eqref{VM-SUSY} and we reproduce them here for convenience.
\begin{align}
  0 = &\; \frac{1}{2}{F_{\mu \nu }}{\sigma ^{\mu \nu }}{\xi _I} + 2{D_\mu }\phi {\sigma ^\mu }{{\tilde \xi }_I} + \phi {\sigma ^\mu }{D_\mu }{{\tilde \xi }_I} + 2i{\xi _I}[\phi ,\tilde \phi ] + {D_{IJ}}{\xi ^J} \label{VMBPSeq1} \\
  0 = &\; \frac{1}{2}{F_{\mu \nu }}{{\tilde \sigma }^{\mu \nu }}{{\tilde \xi }_I} + 2{D_\mu }\tilde \phi {{\tilde \sigma }^\mu }{\xi _I} + \tilde \phi {{\tilde \sigma }^\mu }{D_\mu }{\xi _I} - 2i{{\tilde \xi }_I}[\phi ,\tilde \phi ] + {D_{IJ}}{{\tilde \xi }^J} \label{VMBPSeq2}\;.
\end{align}
From these complex BPS equations, one can derive that $R^\mu F_{\mu \nu} = D_\nu (s \tilde \phi + \tilde s \phi)$, and as a result, $R^\mu D_\mu (s \tilde \phi + \tilde s \phi) = 0$. Moreover, one finds
\begin{align}
  D_{IJ} = & \ -\frac{1}{4s^2} \big[  -2s F_{\mu \nu} + 4(R \wedge d_A \phi)_{\mu \nu} - 4 i (\xi^K \sigma_{\mu \nu} \xi'_K) \phi \big] \Theta_{IJ}^{\mu\nu} \nonumber\\
  =& \ + \frac{1}{4\tilde s^2}\big[  -2 \tilde s F_{\mu \nu} + 4 (R \wedge d_A \tilde \phi)_{\mu \nu} + 4i (\tilde \xi^K \tilde \sigma_{\mu \nu} \tilde \xi'_K) \tilde \phi\big]\widetilde \Theta_{IJ}^{\mu\nu}  \ .\label{DIJ-solution}
\end{align}
Here $\Theta_{IJ}^{\mu\nu} = (\xi_I \sigma^{\mu\nu}\xi_J)$ and similarly $\widetilde\Theta_{IJ}^{\mu\nu} = (\tilde\xi_I \tilde\sigma^{\mu\nu}\tilde\xi_J)$. These relations and expressions will be particularly useful when we evaluate the classical vector and hypermultiplet actions in appendix \ref{app:evaluateCA}.

Let us now impose the reality conditions
\begin{equation}
  {\phi ^\dag } =  - \tilde \phi \Leftrightarrow \phi  = \frac{1}{2}({\phi _2} - i{\phi _1}), \quad   \tilde \phi  = \frac{1}{2}( - {\phi _2} - i{\phi _1}) , \qquad {A^\dag } = A \Leftrightarrow {F^\dag } = F \label{reality-condition-phi-F}\;,
\end{equation}
but keep the auxiliary field $D_{IJ}$ complex. Here $\phi_1,\phi_2$ are real/hermitian fields. The motivation for treating $D_{IJ}$ separately comes from the observation that in fact the auxiliary field $D_{11}(0)$ is a Schur operator. The BPS equations \eqref{VMBPSeq1} and \eqref{VMBPSeq2} can then be decomposed into their real and imaginary parts. We find
\begin{align}
0 = &\; [\phi_1, \phi_2]\;,\\
0 = &\; {D_\mu }(\cos \theta \sin \rho (\cos \varphi {\phi _1} + \sin \varphi {\phi _2}))\;,\quad 0= {D_\varphi }(\cos \theta \sin \rho ( - \sin \varphi {\phi _1} + \cos \varphi {\phi _2})) \;,\label{VMeqn1}\\
F = &\; *\left( {{d_A}\left( {\frac{{r\cos \theta \sin \rho {\phi _2}}}{{\cos \varphi }}d\varphi } \right)} \right)  + *({(s\tilde s)^{ - 1}}R \wedge \mathcal{D}),\label{VM-BPS-equation-F}
\end{align}
as well as
\begin{align}
  \re D_{12} = &  \; - \lambda \sin \theta \tan \frac{\rho }{2}(\im D_{22}\sin \chi  + \re D_{22}\cos \chi ) \\
  \im D_{11} = &  \; - \lambda \tan \frac{\rho }{2} \left(+ 2\sin \chi \sin \theta \im D_{12} +  \lambda \re D_{22}\tan \frac{\rho }{2} \right)\\
  \re D_{11} = &  \; + \lambda \tan \frac{\rho }{2}\left( { - 2\cos \chi \sin \theta \im D_{12} + \lambda \im D_{22}\tan \frac{\rho }{2}} \right)\;.
\end{align}
In a slight abuse of notation, we introduced one-forms $R = R_a e^a$, $R^{IJ} = R^{IJ}_a e^a$, and $\mathcal{D} = D_{IJ}R^{IJ}$. The latter satisfies $R^\mu \mathcal{D}_\mu = 0$. If one further imposes the standard reality condition on the auxiliary field, namely $D_{IJ}^\dagger = - D^{IJ}$, then one can easily see that $\mathcal{D}$ vanishes on the BPS locus. Therefore, the one-form $\mathcal{D}$ captures the deviation of $D_{IJ}$ from the real contour.

The scalars $\phi_{1,2}$ are constrained by the equations in \eqref{VMeqn1}. The left equation implies that
\begin{align}
  \cos \varphi {\phi _1} + \sin \varphi {\phi _2} = \frac{\phi_+}{{\cos \theta \sin \rho }}\;,
\end{align}
in terms of a (covariantly) constant matrix $\phi_+$. The right equation of \eqref{VMeqn1} then becomes
\begin{align}
D_\varphi \Big[ \cos\theta \sin \rho \big( - \frac{\phi_+\sin\varphi}{\cos\theta \sin\rho \cos\varphi} + \frac{\phi_2}{\cos\varphi} \big) \Big] = 0 \ .
\end{align}
For smooth solutions, one should set $\phi_+ = 0$. In appendix \ref{subsec:singular-config}, however, we will show that the singular profile described by $\phi_+$ defines an interesting surface operator. Note also that equation \eqref{VM-BPS-equation-F} can be solved for $\mathcal D$ as
\begin{equation}
\mathcal{D} = \iota _R*F  - {\iota _R}{d_{{A }}}\left( {\frac{{r\cos \theta \sin \rho {\phi _2}}}{{\cos \varphi }}d\varphi } \right)\;.
\end{equation}
Finally, \eqref{VM-BPS-equation-F} also implies $\iota_R F = 0$.

Let us finally discuss a special solution to these equations, where we do impose the standard reality properties on all fields. In particular, we set $\mathcal{D} = 0$. The Bianchi identity applied to \eqref{VM-BPS-equation-F} then states that 
\begin{equation}
  {d_{{A^\Phi }}}{F^\Phi } \propto {d_{{A^\Phi }}}*\Big( {{d_{{A^\Phi }}}\big( {\frac{{r\cos \theta \sin \rho {\phi _2}}}{{\cos \varphi }}d\varphi } \big)} \Big) = 0 \Rightarrow {d_{{A^\Phi }}}\left( {\frac{{r\cos \theta \sin \rho {\phi _2}}}{{\cos \varphi }}d\varphi } \right) = 0\;.
\end{equation}
Plugging this back into the BPS equations for the field strength simply sets $F = 0$. Moreover, the simple fact that $\phi_2$ commutes with itself implies that
\begin{equation}
  {D_{\chi ,\theta ,\rho }}\left( {\frac{{r\cos \theta \sin \rho {\phi _2}}}{{\cos \varphi }}} \right) = 0 \Rightarrow {D_\mu}\left( {\frac{{\cos \theta \sin \rho \phi _2}}{\cos \varphi }} \right) = 0\;.
\end{equation}
Hence the combination ${\cos ^{ - 1}}\varphi (\cos \theta \sin \rho {\phi _2})$ is again covariantly constant, and we have
\begin{equation}
  \cos \theta \sin \rho {\phi _2} = \cos \varphi \phi _ -  \;,
\end{equation}
with another covariantly constant matrix $\phi_-$. For smooth BPS solutions, $\phi_- = 0$, but it is relevant for the defect described in appendix \ref{subsec:singular-config}. In summary, the smooth BPS solutions along the real $D_{IJ}$-contour are given by the trivial configuration
\begin{align}
  {\phi _1} = 0 ,\qquad {\phi _2} = 0, \qquad F = 0, \qquad D_{IJ} = 0\;.
\end{align}
This result is puzzling, since it seems too trivial to reproduce the known result for the four-sphere partition function \cite{Pestun:2007rz}. This provides an additional motivation to relax certain reality properties as we did above.

It is also important to remark, and relatively easy to verify, that the deformation action admits fermionic zero-modes around the bosonic configurations discussed here. As it is beyond the scope of this appendix, we do not present the details of these fermionic zero-modes. The vector multiplet localization computation should take these zero-modes into account properly.

\subsection{Evaluating classical actions}\label{app:evaluateCA}
In section \ref{subsection:classical-action-freeHM}, we evaluated the classical action of a free hypermultiplet on the solutions of the complex BPS equations described by \eqref{freeHM-BPS-equation-without-reality}. In this appendix, we present the generalization of that computation to hypermultiplets coupled to a dynamical vector multiplet. We find that both the vector multiplet and gauged hypermultiplet action reduce to a two-dimensional action.

The relevant complex BPS equations for the gauged hypermultiplet are
\begin{align}
  D_\varphi q_{IA} = & - \frac{ir}{2\kappa_0} (s \tilde \phi + \tilde s \phi){_A}^B q_{IB}, \label{fund-HM-BPS-equations-q} \\
  {F_{JA}} = &  - \frac{1}{{\tilde s}}\left[ {2({{\hat \xi }_{I'}}{\sigma ^\mu }{{\tilde \xi }_I}){D_\mu }{q^I}_A + ({{\hat \xi }_{I'}}{\sigma ^\mu }{D_\mu }{{\tilde \xi }_I}){q^I}_A + 4i({{\hat \xi }_{I'}}{\xi _I})\tilde \phi {^A}{_B}{q^{IB}}} \right] \label{fund-HM-BPS-equations-F}\\
  = &  + \frac{1}{s}\left[ {2({{\tilde {\hat \xi} }_{I'}}{{\tilde \sigma }^\mu }{\xi _I}){D_\mu }{q^I}_A + ({{\tilde {\hat \xi} }_{I'}}{{\tilde \sigma }^\mu }{D_\mu }{\xi _I}){q^I}_A + 4i({{\tilde {\hat \xi} }_{I'}}{{\tilde \xi }_I}){\phi ^A}_B{q^{IB}}} \right] \nn\;.
\end{align}
Using these equations together with the complex BPS equations expressing the auxiliary field $D_{IJ}$ given in \eqref{DIJ-solution}, the bosonic part of the hypermultiplet action \eqref{HMaction} can be shown to become
\begin{equation}
S_\text{HM} = \int d^4x \sqrt{g} \nabla_\mu \Big[ \frac{1}{s \tilde s} \epsilon^{\mu \nu \lambda \delta} R_\lambda R^{IJ}_\delta D_\nu q_J^Aq_{IA}  - \frac{2i}{\tilde s} (\tilde \xi_K \tilde \sigma^\mu \xi'_I)q^{KA}q^{I}_A + \frac{i}{s \tilde s} (s\tilde \phi - \tilde s \phi)_{AB}R_{IJ}^\mu q^{IA}q^{JB} \Bigg] \; ,
\end{equation}
while the bosonic part of the vector multiplet action \eqref{YMaction} reduces to
\begin{equation}
S_\text{VM} = \int d^4x \sqrt{g} \nabla_\mu  \tr\Big[ -8i \tilde s^{-1} \tilde K^\mu \phi \tilde \phi - (s \tilde s)^{-1}\epsilon^{\mu \nu \alpha \beta} R_\nu F_{\alpha \beta}(s \tilde \phi - \tilde s \phi)  \Big] \;.
\end{equation}
In the latter we defined $\tilde K^\mu \equiv (\xi'^I\sigma^\mu \tilde \xi_I)$.

Let us first focus on $S_\text{HM}$. The BPS equation \eqref{fund-HM-BPS-equations-q} implies that gauge invariant combinations of $q_{IA}$ are independent of the coordinate $\varphi$. Therefore the total derivative $\partial_\mu (\sqrt{g} [\ldots]^\mu)$ in the above action is in fact a three-dimensional total derivative $\partial_{\bar\mu} (\sqrt{g} [\ldots]^{\bar\mu})$, with $\bar \mu=\chi,\theta,\rho$. At this point we can perform the $\varphi$-integral. Also notice that the index $\nu$ in $D_\nu q^A_J$ inside the bracket can never take the value $\varphi$. We are left with the hypermultiplet action as a three-dimensional integral. By Stokes' theorem, it reduces to a two-dimensional action on $S^2_{\theta = \frac{\pi}{2}}$ 
\begin{align}
S_\text{HM}^\text{2d}[\mathfrak{q}] = - 2\pi r\int \sqrt {{g_{{S^2}}}} d\rho d\chi \bigg[ \Omega^{AB} \epsilon ^{\mu\nu} ({\mathfrak{q}^\text{2d}_A}{\gamma _\mu}{D^\text{2d}_\nu}\mathfrak{q}^\text{2d}_B) + & \left .\frac{{i{{(s\tilde \phi  - \tilde s\phi )}_{AB}}R_{IJ}^3}}{{4\kappa^2{\lambda ^2}\cos \theta \sin \rho }}{q^{IA}}{q^{JB}} \right |_{\theta \to \pi/2} \bigg]\;, \nn
\end{align}
where $D^\text{2d}_\mu$ contains the BPS vector multiplet gauge field $A_\mu$. It acts on the indices $A,B = 1, \ldots, 2N$ in the appropriate way. For smooth $\phi$ and $\tilde \phi$, the second term goes to zero in the $\theta\rightarrow \frac{\pi}{2}$ limit. Similarly to section \ref{subsection:classical-action-freeHM}, the action can finally be written more explicitly as
\begin{equation}\label{gaugedHMaction2d}
S_{{\text{HM}}}^{{\text{2d}}}[Q] = - 4\pi i\int d\chi d\rho \sqrt {g_{S^2}}\ \widetilde Q\; \overline{\wp}_A \; Q\;,
\end{equation}
where $\overline\wp_A$ is the gauged monopole-charge-lowering operator of monopole charge $\mathfrak m=\frac{1}{2}$.

Let us now consider the vector multiplet action $S_{\text{VM}}$. Recall from around \eqref{DIJ-solution} that
\begin{align}
  R^\nu F_{\mu \nu} + D_\mu (\tilde s\phi + s \tilde \phi) = 0, \quad R^\mu D_\mu (s \tilde \phi) = - R^\mu D_\mu (\tilde s \phi) = i [\tilde s\phi, s \tilde \phi]\; .
\end{align}
As a result, up to a gauge transformation, $s\tilde \phi$, $\tilde s \phi$ and $F$ are independent of $\varphi$. We can thus first perform the $\varphi$ integration. Then completely similarly to the analysis of the hypermultiplet, the action can be written as a total derivative to which we can apply Stokes' theorem. We end up with an action on $S^2_{\theta=\frac{\pi}{2}}$
\begin{equation}
S_\text{VM} = 2\pi \int d\chi d\rho \sqrt {g_{S^2}}\tr\left[\left. - 4\phi \tilde \phi  + \frac{1}{r \sin ^2\rho }{F_{\chi \rho }}(s\tilde \phi  - \tilde s\phi )\right|_{\theta \to \pi/2}\right] \; .
\end{equation}
One can verify that for smooth $\phi$, $\tilde \phi$, the second term vanishes and we have
\begin{equation}
S_\text{VM} = - 8 \pi \int d\chi d\rho \sqrt {g_{S^2}}\left.\tr(\phi \tilde \phi)\right|_{\theta \to \pi/2} \ .
\end{equation}

In view of the fact that the vector multiplet possesses fermionic zero-modes, it is of interest to also consider the fermionic part of the super Yang-Mills action. We will show here that it also reduces to a two-dimensional action. For simplicity, we consider the free limit, where $g_\text{YM} = 0$. In this case, the fermionic part consists of the Dirac action,
\begin{equation}
 S_\text{Dirac} \propto \int d^4 x \sqrt{g_{S^4} } \tr ( \lambda^I \sigma^\mu \nabla_\mu \tilde \lambda_i) \; .
\end{equation}
Let us define $\Lambda^\mu \equiv (\lambda^I \sigma^\mu \tilde \xi_I)$, $\tilde \Lambda^\mu \equiv (\xi ^I \sigma^\mu \tilde \lambda^I )$. The BPS equations impose various conditions on $\Lambda$ and $\tilde \Lambda$, including \footnote{We have used the BPS equations as they follow from the gauge-fixed vector multiplet, hence the appearance of the $c$-ghost. While in intermediate steps the ghost will appear in various places, in the final result it drops out.}
\begin{equation}
R^\mu \Lambda_\mu = 0, \quad R^\mu \tilde \Lambda_\mu = 0, \quad \tilde \Lambda = - \Lambda_\mu + i \partial_\mu c\ , \quad \mathcal{L}_R \Lambda = 0, \quad R^\mu \partial_\mu c = 0\ .
\end{equation}
Using the identities
\begin{align}
  {\lambda ^I} = \frac{1}{{\tilde s}}({\lambda ^L}{\sigma ^\lambda }{{\tilde \xi }_L}){\sigma _\lambda }{{\tilde \xi }^I},\qquad {{\tilde \lambda }_I} = \frac{1}{s}({\xi ^K}{\sigma ^\mu }{{\tilde \lambda }_K}){{\tilde \sigma }_\mu }{\xi _I}\;,
\end{align}
one can straightforwardly derive that
\begin{equation}
S_\text{Dirac} \propto \int d^4 x \sqrt{g_{S^4}} \nabla_\mu \tr (\epsilon^{\mu\nu \lambda\rho} (s \tilde s)^{-1}R_\lambda  \Lambda_\nu\Lambda_\rho) \; ,
\end{equation}
which reduces to
\begin{align}
  S_\text{Dirac} \propto 2\pi r\int d\rho d\chi \sqrt{g_{S^2}} \tr \Lambda_2 \Lambda_4 \ .
\end{align}

\subsection{Some Gaussian integrals}\label{appendix_oneloop}
Let us define an interesting BPS background configuration for the vector multiplet. We take the auxiliary fields $D_{IJ}$ to be complex, so that in fact there exist nontrivial, smooth BPS configurations. We choose the gauge group to be $SU(N)$ and take $A_\mu$ at $S^2_{\theta=\frac{\pi}{2}}$ to have an $\mathfrak{su}(N)$ Cartan-valued monopole-like profile,
\begin{equation}
A|_{S^2_{\theta = \pi/2}}\sim \operatorname{diag}({\mathfrak{a} _1},...,{\mathfrak{a} _N})(\cos \rho  - 1)d\chi \;,\qquad \sum_{a = 1}^N \mathfrak{a}_a = 0\;.\label{monoopole-background}
\end{equation}
We recall that any $SU(N)$ bundle over $S^2$ is trivial as $\pi_1 SU(N) = \{e\}$. Therefore, unlike $U(1)$ monopoles, there is no quantization condition on $\mathfrak{a}$. The fact that the $SU(N)$-bundle over $S^2_{\theta = \frac{\pi}{2}}$ is trivial also implies that one can extend $A_\mu$ into the interior of $S^2_{\theta = \frac{\pi}{2}}$ and therefore to the entire four-sphere.

Let us consider fundamental hypermultiplets coupled to this vector multiplet background. The relevant two-dimensional theory is described by the action \eqref{gaugedHMaction2d}. We would like to compute the resulting partition function. Following the discussion in subsection \ref{subsec: evaluation oneloop}, we expand the components of $Q$ and $\widetilde Q$ in the basis of harmonics $Y^{1/2}_{jm}$ and $\overline{Y^{-1/2}_{jm}}$ respectively, with complex conjugate coefficients $Q_{ajm}$ and $\overline{Q_{ajm}}$. Here $a = 1, \ldots, N$ is the gauge index. Using the orthonormality properties of $Y_{jm}^\mathfrak{m}$, the action simplifies to
\begin{small}
\begin{equation}\label{gI}
S_\text{HM}^\text{2d} = 4\pi i\sum\limits_{a = 1}^N {\sum\limits_{j',j = 1/2}^{ + \infty  + 1/2} {\sum\limits_{m' =  - j'}^{ + j'} {\sum\limits_{m =  - j}^{ + j} {\overline {{Q_{aj'm'}}} \left[ {(j + 1/2){\delta _{mm'}}{\delta _{jj'}} - {\mathfrak{a}_a}\ {}_{\frac{1}{2}}\langle j'm'|\frac{{\cos \rho  - 1}}{{\sin \rho }}|jm\rangle_{\frac{1}{2}} } \right]{Q_{ajm}}} } } }  \;,
\end{equation}
\end{small}%
where the second term in the square bracket is given by\footnote{We checked the final equality with Mathematica systematically to very large $j$ and $j'$.}
\begin{multline}
{}_{\frac{1}{2}}\langle j'm'| \frac{{\cos \rho  - 1}}{{\sin \rho }}|jm\rangle_{\frac{1}{2}} \equiv  \int \sqrt{g_{S^2}} d\rho d\chi\ \overline{Y^{-1/2}_{j'm'}(\rho, \chi)}\;\frac{\cos\rho - 1}{\sin \rho}\; Y^{1/2}_{jm}(\rho, \chi)  \label{def:matrix-element}\\
  = \ {\delta _{mm'}}\frac{m}{{|m|}}\left[ { - {\delta _{jj'}} + 2({\delta _{j' > j}}{\delta _{m < 0}} + {\delta _{j' < j}}{\delta _{m > 0}}){{( - 1)}^{j' + j}}\sqrt {\frac{{{{( - {{\min }_{j,j'}} - |m|)}_{2|m|}}}}{{{{( - {{\max }_{j,j'}} - |m|)}_{2|m|}}}}} } \right] \nn\;.
\end{multline}
Here we used the Pochhammer symbol $(x)_n \equiv \prod_{k=0}^{n-1}(x+k)$, and the delta function $\delta_p = 1$ if the proposition $p$ is true and $\delta_p = 0$ otherwise.

Using this result, one can perform the Gaussian integral in \eqref{gI} explicitly, and one finds up to some unimportant constant prefactors,
\begin{align}
  Z_\text{HM} = \lim_{b \to 1} \prod_{a=1}^N\Upsilon_b \Big(\frac{b + b^{-1}}{2} + \mathfrak{a}_a \Big) \;.
\end{align} 
Quite curiously, this is precisely the one-loop determinant for a gauged hypermultiplet. Quite possibly, this result implies that the monopole-configuration we turned on plays an essential role in the full localization computation including dynamical vector multiplets. 

As reviewed in section \ref{section:review-chiral-algebra}, the chiral algebra associated with the free vector multiplet is a small $(b,c)$ ghost system. Without derivation, but given our experience of section \ref{section:freeHM}, we consider the following action on $S_{\theta = \frac{\pi}{2}}^2$
\begin{equation}
S_{{\text{VM}}}^{{\text{2d}}} = \int {\sqrt {{g_{{S^2}}}} d\rho d\chi\ \tr b\left( {{\partial _\rho } - \frac{i}{{\sin \rho }}{\partial _\chi }} \right)c} \;.
\end{equation}
Here $b$ and $c$ are fermionic scalars in the adjoint representation. Let us perform a simple check by comparing its partition function to the partition function of a four-dimensional free vector multiplet. We expand $c$ and $b$ as
\begin{equation}
b = \sum_{j = 1}^\infty  {\sum_{m =  - j}^{ + j} {{b_{jm}} \overline{Y_{jm}^{s = 1}}} } ,\qquad c = \sum_{j = 0}^\infty  {\sum_{m =  - j}^{ + j} {{c_{jm}}Y_{jm}^{s = 0}} }  = {c_0}Y^{s=0}_{00} + \sum_{j = 1}^\infty  {\sum_{m =  - j}^{ + j} {{c_{jm}}Y_{jm}^{s = 0}} }  \;,
\end{equation}
and assume the reality condition $\overline{b_{jm}} = c_{jm}$. Note that we separated out the $c$ zero-mode. This mode is absent in the small $(b,c)$ ghost system. Each root contributes to the partition function as
\begin{align}
\prod\limits_{j = 1}^\infty  {{{\sqrt {j(j + 1)} }^{2j + 1}}}  = & \prod\limits_{j = 1}^\infty  {{{\sqrt j }^{2j + 1}}} \prod\limits_{j = 1}^\infty  {{{\sqrt {j + 1} }^{2j + 1}}}  \\
  = & \prod\limits_{j = 1}^\infty  {{{(j + 1)}^{2(j + 1)}}} = \prod\limits_{m,n \geqslant 0} {(m + n + 2)(m + n)'} \\
  = &\; {\lim _{x \to 0}}\frac{{\Upsilon {{(x)}_{b = 1}}}}{x}\;.
\end{align}
Raising this result to the power of the dimension of the gauge group, we indeed recover the partition function of a free vector multiplet.

Let us finally consider the small $(b,c)$ ghost system in the presence of the background defined in \eqref{monoopole-background}. The action on $S^2_{\theta = \pi/2}$ then reads
\begin{equation}
S_{{\text{VM}}}^{{\text{2d}}} = \int {\sqrt {{g_{{S^2}}}} d\rho d\chi\ \tr b\left( {{\partial _\rho } - \frac{i}{{\sin \rho }}({\partial _\chi } - i{A_\chi })} \right)c} \;,
\end{equation}
We again expand the components of $b,c$ in terms of $\overline{Y_{jm}^{\mathfrak{q} = -1}}$ and $Y_{jm}^{\mathfrak{q} = 0}$. Representing the adjoint $SU(N)$ index in terms of a fundamental and antifundamental index, we obtain
\begin{equation}
S_{{\text{VM}}}^{{\text{2d}}} = \sum_{a,b} {\sum_{j' = 1}^{ + \infty } {\sum_{j = 1}^{ + \infty } {\sum_{m' =  - j'}^{ + j'} {\sum_{m =  - j}^{ + j} {\overline {(c^a_b)_{j'm'}} \left( {\sqrt {j(j + 1)}  - ({\mathfrak{a}_a} - {\mathfrak{a}_b})\ {}_{1}\langle j'm'|\frac{{\cos \rho  - 1}}{{\sin \rho }}|jm\rangle_0 } \right)(c^b_a)_{jm}} } } } }\;, 
\end{equation}
where
\begin{equation}
{}_{1}\langle j'm'|\frac{{\cos \rho  - 1}}{{\sin \rho }}|jm\rangle_0  = \int {\sqrt {{g_{{S^2}}}} d\rho d\chi\ \overline {Y_{j'm'}^{\mathfrak{q} =  - 1}} \;\frac{{\cos \rho  - 1}}{{\sin \rho }}\;Y_{jm}^{\mathfrak{q} = 0}} \;.
\end{equation}
Finally, performing the Gaussian integrals, we find
\begin{equation}
  {Z_{{\text{VM}}}}(\mathfrak{a}) = \prod\limits_{a > b} {\frac{{\Upsilon ({\mathfrak{a}_a} - {\mathfrak{a}_b})\Upsilon ( - {\mathfrak{a}_a} + {\mathfrak{a}_b})}}{{{{({\mathfrak{a}_a} - {\mathfrak{a}_b})}^2}}}}  = Z_{{\text{1-loop-vector}}}^{{S^4}}(\mathfrak{a})\;,
\end{equation}
up to some numerical prefactors.

\subsection{\texorpdfstring{$(0,4)$}{(0,4)}-supersymmetric singular configurations}\label{subsec:singular-config}
In appendix \ref{appendix_VMBPS} we encountered interesting singular solutions to the vector multiplet BPS equations. In this appendix we would like to show that they define a surface defect that preserves two-dimensional $(0,4)$ superconformal symmetry. We do so by finding all conformal Killing spinors under which the profile is supersymmetric.

The singular profile for the vector multiplet scalars is paramterized by two commuting constant matrices $\phi_+$, $\phi_-$,
\begin{equation}
\phi = \frac{1}{2 \sin\rho \cos \theta}e^{ + i \varphi} (\phi_+ - i \phi_-)\;, \qquad \tilde\phi = - \frac{1}{2 \sin\rho \cos \theta}e^{ - i \varphi} (\phi_+ - i \phi_-) \;,
\end{equation}
and the field strength and auxiliary field are set to zero.\footnote{Note that the space $S^4 - S^2_{\theta = \frac{\pi}{2}}$ is not simply-connected, and therefore flat connections can have nontrivial holonomy around the sphere $S^2_{\theta = \frac{\pi}{2}}$.}

It is easy to verify that an eight-parameter family of conformal Killing spinors preserves the above singular profile. Concretely,
\begin{align}
  & {\xi _1} = \alpha_1^{-1,1}\cos \frac{\rho }{2}\left( {\begin{array}{*{20}{c}}
    {{e^{\frac{i}{2}( + \theta  + \varphi  - \chi )}}} \\ 
    {{e^{\frac{i}{2}( - \theta  + \varphi  - \chi )}}} 
  \end{array}} \right) + \alpha_1^{11}\sin \frac{\rho }{2}\left( {\begin{array}{*{20}{c}}
    {{e^{\frac{i}{2}( - \theta  + \varphi  + \chi )}}} \\ 
    { - {e^{\frac{i}{2}( + \theta  + \varphi  + \chi )}}} 
  \end{array}} \right) \nonumber\\
  & {\xi _2} = \alpha_2^{-1,1}\cos \frac{\rho }{2}\left( {\begin{array}{*{20}{c}}
    {{e^{\frac{i}{2}( + \theta  + \varphi  - \chi )}}} \\ 
    {{e^{\frac{i}{2}( - \theta  + \varphi  - \chi )}}} 
  \end{array}} \right) + \alpha_2^{1,1}\sin \frac{\rho }{2}\left( {\begin{array}{*{20}{c}}
    {{e^{\frac{i}{2}( - \theta  + \varphi  + \chi )}}} \\ 
    { - {e^{\frac{i}{2}( + \theta  + \varphi  + \chi )}}} 
  \end{array}} \right) \\
  & {{\tilde \xi }_1} = \tilde \alpha_1^{-1,-1}\cos \frac{\rho }{2}\left( {\begin{array}{*{20}{c}}
    {{e^{\frac{i}{2}( - \theta  - \varphi  - \chi )}}} \\ 
    {{e^{\frac{i}{2}(+ \theta  - \varphi  - \chi )}}} 
  \end{array}} \right) + \tilde \alpha_1^{1,-1}\sin \frac{\rho }{2}\left( {\begin{array}{*{20}{c}}
    {{e^{\frac{i}{2}(+ \theta  - \varphi  + \chi )}}} \\ 
    { - {e^{\frac{i}{2}( - \theta  - \varphi  + \chi )}}} 
  \end{array}} \right) \nonumber\\
  & {{\tilde \xi }_2} = \tilde\alpha_2^{-1,-1}\cos \frac{\rho }{2}\left( {\begin{array}{*{20}{c}}
    {{e^{\frac{i}{2}( - \theta  - \varphi  - \chi )}}} \\ 
    {{e^{\frac{i}{2}(+ \theta  - \varphi  - \chi )}}} 
  \end{array}} \right) + \tilde\alpha_2^{1,-1}\sin \frac{\rho }{2}\left( {\begin{array}{*{20}{c}}
    {{e^{\frac{i}{2}(+ \theta  - \varphi  + \chi )}}} \\ 
    { - {e^{\frac{i}{2}( - \theta  - \varphi  + \chi )}}} 
  \end{array}} \right)\ , \nonumber
\end{align}
with constants $\alpha_I^{\pm 1, \pm 1}$ and $\tilde\alpha_I^{\pm 1,\pm 1}$. The subalgebra generated by the corresponding supercharges is precisely the (centrally extended) two-dimensional $\mathcal N=(0,4)$ superalgebra on $S^2_{\theta=\frac{\pi}{2}}$.

\clearpage

{
\bibliographystyle{utphys}
\bibliography{ref}
}

\end{document}